\documentclass[twocolumn]{aastex631}

\usepackage{graphicx,caption,subcaption}
\usepackage{array}
\usepackage{amsmath}

\newcommand{\gdone}{GD-1}

\begin{document}
\title{Towards characterizing dark matter subhalo perturbations in stellar streams with graph neural networks}

\author[0000-0001-8975-3719]{Peter Xiangyuan Ma}
\affiliation{Department of Astronomy, UC Berkeley, 501 Campbell Hall,
Berkeley, CA, 94720, United States of America}
\affiliation{Department of Mathematics, University of Toronto, 40 St. George Street, Toronto, ON, M5S 2E4, Canada}

\correspondingauthor{Peter Xiangyuan Ma}
\email{peter\_ma@berkeley.edu}

\author{Keir K. Rogers}
\affiliation{Department of Physics, Imperial College London, Blackett Laboratory, Prince Consort Road, London, SW7 2AZ, United Kingdom}
\affiliation{Dunlap Institute for Astronomy and Astrophysics, University of Toronto, 50 St. George Street,
Toronto, ON, M5S 3H4, Canada}

\author[0000-0002-9110-6163]{Ting S. Li}
\affiliation{David A. Dunlap Department of Astronomy and Astrophysics, University of Toronto,\\ 50 St. George Street,
Toronto, ON, M5S 3H4, Canada}

\author{Ren\'ee Hlo\v{z}ek}
\affiliation{Dunlap Institute for Astronomy and Astrophysics, University of Toronto, 50 St. George Street,
Toronto, ON, M5S 3H4, Canada}
\affiliation{David A. Dunlap Department of Astronomy and Astrophysics, University of Toronto,\\ 50 St. George Street,
Toronto, ON, M5S 3H4, Canada}

\author{Jeremy J. Webb}
\affiliation{Department of Science, Technology and Society, Division of Natural Science, York University,\\ 218 Bethune College, Toronto, ON, M3J 1P3, Canada}
\affiliation{David A. Dunlap Department of Astronomy and Astrophysics, University of Toronto,\\ 50 St. George Street,
Toronto, ON, M5S 3H4, Canada}

\author{Ruth Huang}
\affiliation{David A. Dunlap Department of Astronomy and Astrophysics, University of Toronto,\\ 50 St. George Street,
Toronto, ON, M5S 3H4, Canada}

\author{Julian Meunier}
\affiliation{David A. Dunlap Department of Astronomy and Astrophysics, University of Toronto,\\ 50 St. George Street,
Toronto, ON, M5S 3H4, Canada}

\begin{abstract}
The phase space of stellar streams is proposed to detect dark substructure in the Milky Way through the perturbations created by passing subhalos — and thus is a powerful test of the cold dark matter paradigm and its alternatives. Using graph convolutional neural network (GCNN) data compression and simulation-based inference (SBI) on a simulated GD-1-like stream, we improve the constraint on the mass of a [$10^8$, $10^7$, $10^6$] $M_\odot$ perturbing subhalo by factors of [11, 7, 3] with respect to the current state-of-the-art density power spectrum analysis. We find that the GCNN produces posteriors that are more accurate (better calibrated) than the power spectrum. We simulate the positions and velocities of stars in a GD-1-like stream and perturb the stream with subhalos of varying mass and velocity. Leveraging the feature encoding of the GCNN to compress the input phase space data, we then use SBI to estimate the joint posterior of the subhalo mass and velocity. We investigate how our results scale with the size of the GCNN, the coordinate system of the input and the effect of incomplete observations. Our results suggest that a survey with $10 \times$ fewer stars (300 stars) with complete 6-D phase space data performs about as well as a deeper survey (3000 stars) with only 3-D data (photometry, spectroscopy). The stronger constraining power and more accurate posterior estimation motivate further development of GCNNs in combining future photometric, spectroscopic and astrometric stream observations.

\end{abstract}


\section{Introduction} \label{sec:intro}
The cold dark matter (CDM) model remains most preferred given astrophysical and cosmological observations \citep{plank2018}, but its fundamental nature is undetermined despite an extensive direct detection program \citep[e.g.,][]{Cooley:2022ufh}. The hierarchical clustering within the CDM paradigm is successful in matching theory with observations of structure formation from the largest scales (\(\sim\) Gpc) down to galactic scales (\(\sim\) Mpc): dark matter (DM) forms halos from galaxy cluster masses (\(\sim 10^{15} M_\odot\)) down to dwarf galaxy masses (\(\sim 10^{8} M_\odot\)). The CDM model also predicts that substructure will form on even smaller scales, i.e., in subhalos with masses \(< 10^{8} M_\odot\) that do not host any baryonic matter. However, this regime remains largely untested. It is therefore a focus of current and upcoming photometric and spectroscopic surveys, e.g., the Vera C. Rubin Observatory \citep[\textit{Rubin},][]{lsstdm_review}, \textit{Euclid} \citep{euclid}, the Dark Energy Spectroscopic Instrument \citep[DESI,][]{DESI_whitepaper} and the Nancy Grace Roman Space Telescope \citep[\textit{Roman},][]{roman}, to infer robustly the existence of dark substructure within the Milky Way (MW). Doing so would be a powerful confirmation of the CDM paradigm. On the other hand, proving the absence (or, indeed, enhancement) of substructure would help distinguish between theoretically well-motivated alternatives to CDM.

Although subhalos are not directly observable, there are indirect probes of their presence. Strong gravitational lenses are sensitive to substructure along the line of sight from source to observer, both within and outside the lens \citep{Mandelbaum2006,lense,Vegetti2010,lensing_review}, with current sensitivity down to $\sim (10^7 - 10^8) M_\odot$. The Lyman-$\alpha$ forest, a spectral feature formed in the intergalactic medium (IGM), traces quasi-linear DM fluctuations (in the filaments and voids where the IGM resides) on the smallest scales currently accessible \citep{Croft1999,Croft2002,McDonald2000, lyman, Boera2019,Villasenor2023}. \cite{lyman_low, Rogers:2020cup, keir_peiris_dvorkin} set a limit on the allowed minimum half-mode halo mass of \(7.2\,\times\,10^7\,M_\odot\) (95 \% c.l.). The luminosity function of MW satellite galaxies can be related to the mass function of the subhalos that host them \citep{Vale2004, Macci2010, Wechsler2018, satellites}. Combined with strong lensing, the minimum half-mode halo mass (for a warm DM transfer function) is limited above $\sim 10^7 M_\odot$ \citep[95 \% c.l.,][]{satellites_lensing}. One of the most powerful probes of substructure in upcoming surveys is the phase space of streams of stars in the Milky Way, e.g., the Legacy Survey of Space and Time (LSST) from \textit{Rubin} is forecast to probe down to \(\sim 10^5 M_\odot\) \citep{lsstdm_review}.

\subsection{Stellar streams as a probe of dark substructure}
Stellar streams are long, thin and dynamically cold trails of stars \citep{stream_ghost} formed from tidally disrupted globular clusters and dwarf galaxies \citep{distrupted}. The formation of a stream in the absence of DM substructure can be modeled (see \S~\ref{sec:intro_sims}) by accounting for the dynamical tidal disruption \citep{dynamical}, the time-varying background potential of the MW \citep{bar}, merger with the Large Magellanic Cloud and interaction with baryonic (as opposed to DM) substructure like giant molecular clouds \citep{lmc}. Any additional perturbations in the structure of the stream (both star positions and kinematics) will indicate interactions with DM subhalos \citep{Johnston2002, 2002MNRAS.332..915I, Erkal2015, stream_gap, stream_dm_jo}. Dozens of streams have been discovered and characterized in the MW halo, thanks to deep- and wide-field photometric and spectroscopic surveys \citep[e.g.,][]{Koposov2014,Shipp2018,Jethwa2018, Li2019}, as well as astrometric data provided by \textit{Gaia} \citep[e.g.,][]{Ibata2019, Ibata2021,Malhan2024}. Several stellar streams have already been studied for evidence of density variations and potential signatures of dark halo interactions. Notable examples include GD-1 \citep{gd-1, PriceWhelan2018, stream}, Palomar 5 \citep{Odenkirchen2001, Carlberg2012, Erkal2017} and Atlas-Aliqa Uma \citep{Koposov2014, atlas, Hilmi2024}, among others \citep[e.g.,][]{Patrick2022,Tavangar2022}.

The current state-of-the-art in the full statistical analysis of {\gdone} (i.e., beyond modeling specific, identifiable features) is to measure the 1D angular power spectrum of the star density along the arc of the stream, probing subhalo masses from $10^6 M_\odot$ to $10^8 M_\odot$ \citep{2017MNRAS.466..628B,warm_dm_sbi,Banik2021_obs}.

It is manifest that the 1D angular power spectrum does not extract all the information contained in the stream. First, information in the second angular coordinate is lost as it is known that streams contain structures away from the main orbit like spurs of stars, which could be a clear signature of a subhalo perturbation \citep{stream_gap}. Second, by combining photometry with astrometry and spectroscopy, the full 6-D phase space of the stream can be built up (at least for a subset of stars observed in a given stream). E.g., the additional kinematic data from \textit{Gaia} \citep{gaia} provide two velocity components. \textit{Rubin} LSST is expected to measure proper motions over a decade-long baseline at the current \textit{Gaia} precision but with images that are three magnitudes deeper \citep{stream_obs}. Spectroscopic measurements, e.g., from the Southern Stellar Stream Spectroscopy Survey \citep[$S^5$,][]{Li2019} or DESI \citep{Cooper2023}, also provide line-of-sight velocities helping to complete the 6-D phase space. It is therefore necessary to prepare future analysis pipelines to leverage the increasingly rich data that we anticipate from deep and wide photometric and spectroscopic surveys.

\subsection{Machine learning and graph neural networks}
\label{sec:graph_intro}
The challenge in using streams to constrain DM is to infer the model parameters of a subhalo (or subhalos) interacting with a stream (in particular, the subhalo mass) given the (incomplete) 6-D phase space of hundreds or thousands of observable stars in a stream. As discussed above, the 1D angular power spectrum is a lossy compression of the data. However, field-level inference, where inference occurs directly from the positional data (although some form of neural compression is, in practice, often necessary), is increasingly demonstrated to be a viable and powerful approach, e.g., for the cosmic microwave background \citep{Caldeira2019}, galaxy clustering \citep{Lemos2024}, the Lyman-\(\alpha\) forest \citep{Nayak2024}, astrometric lensing \citep{MishraSharma2022}. In these examples, data are compressed using a convolutional neural network. \cite{field_minh} directly infers from the galaxy field without compression.

Given that we want to combine photometric, astrometric and spectroscopic observations, our data are better described as a point cloud, where each point (star) is labeled with a 6-D phase space vector. For such data, graph convolutional neural networks \citep[GCNNs,][]{GCNN} are a suitable compression. GCNNs exploit the graph-like geometry that arises in modeling point cloud systems (usually from the spatial arrangement of these points). \textcolor{black}{GCNNs are superior in our setting because point clouds are inherently irregular and unordered, making traditional neural networks less effective. Further, GCNNs capture both local and global features which is critical for leveraging the geometric structure of the cloud. Image convolutions are not appropriate because the raw data are a mix of images, spectra and time series, which we instead represent as features on the graph.} The graph structure is formed by identifying graph nodes as stars and graph edges as connections between neighbouring stars. The graph can thus capture pairwise interactions present in our data. GCNNs are successfully demonstrated to infer DM density profiles from star data in dwarf galaxies \citep{gnn_dm_streams}, to infer baryonic properties from dark matter subhalo properties \citep{gcnn_baryons}, to carry out symbolic regression from DM-only simulations \citep{graph_dm}. In this work, we investigate using GCNNs to compress the phase space of a stream down to estimators of the perturbing subhalo mass and velocity.

\subsection{Simulation-based inference}
After this neural compression of the data, standard approaches to parameter inference are typically no longer feasible, as the likelihood function can not now be easily analytically formed. However, since we can simulate the stream, we can therefore forward model the data and learn the posterior or likelihood directly from samples of the joint distribution of data and parameters. This concept encompasses a broad class of algorithms called simulation-based inference \citep[SBI,][]{Diggle1984, likelihood-free, norm_flow}. Within this class, machine learning models, e.g., Gaussian mixture models \citep{gaussian_mixture_models} or normalizing flows \citep{norm_flows_1,norm_flow_2, norm_flow}, learn either the posterior \citep{posterior_est_1, posterior_est_2, posterior_est_3}, the likelihood \citep{likelihood_est} or the ratio of likelihood to evidence \citep{ratio_est, ratio_est_2}. SBI is increasingly used in, e.g., cosmology given the need to model complex and multivariate likelihoods \citep[e.g.,][]{bofi, delfi, sbi_cmb,Chen:2023cfb,Lemos2023,Lin2023}. SBI is also successfully applied in the analysis of stellar streams. E.g., \cite{warm_dm_sbi} uses SBI to infer the warm DM particle mass given stream density variations; \cite{stream_sbi} uses SBI given the full phase space of a stream; \cite{galactic_acceleration} uses SBI to reconstruct the galactic acceleration field. Here, we use neural posterior estimation with a normalizing flow model \citep{sbi-toolkit} to learn the posterior distribution, using the GCNN as a data compression.

\subsection{Stellar stream simulations}
\label{sec:intro_sims}
Since the GCNN and SBI models require \(\mathcal{O} (10^3 - 10^4)\) training simulations, we must simulate stellar streams with a balance of computational efficiency and physical accuracy that gives robust inference. The most accurate, but most computationally expensive, approach is a hydrodynamical simulation of the Milky Way, e.g., the Feedback in Realistic Environments \citep[FIRE,][]{fire, Shipp2023} simulations or the Numerical Investigation of Hundred Astrophysical Objects \citep[NIHAO,][]{nihao, nihao2} simulations. These simulations assume a fluid model of the formation of galaxies and their satellites, streams and substructure. They incorporate baryonic feedback (both mechanical and radiative) sources that also shape the DM distribution, e.g., stellar and supernovae feedback, feedback from active galactic nuclei and black holes, star formation quenching. These codes are computationally expensive: a lower-resolution \textsc{FIRE} simulation with $10^6$ particles requires $> 10$ hours \citep{fire}. There are faster \(N\)-body simulations of dissolving star clusters \citep[e.g.,][]{nbody1, nbody2, nbody3, nbody4} that neglect detailed hydrodynamics, but model internal cluster evolution with much higher precision. It is still intractable to run these simulations in sufficient number for the training of our machine learning models.

Best suited for our purposes are particle spray-based methods \citep[e.g.,][]{streamspray_original, chen24} that generate realistic representations of stellar streams without the need for a detailed cluster or galaxy simulation. We use the \textsc{streamspraydf} algorithm \citep{streamspray_original} as implemented in the \textsc{galpy} Galactic dynamics package \citep{galpy, Qian2022}. We model a single stream by randomly ejecting stars from a progenitor. In this proof-of-principle work, we consider interactions between a {\gdone}-like stream and a single subhalo, although a real stream will appreciably interact with \(\mathcal{O}(100)\) subhalos \citep{Diemand2007,Diemand2008,stream_gap}. This approach is computationally tractable to run in large numbers and is demonstrated to reproduce observed stream properties like their extent and the ``streaky'' features arising from the disruption of the progenitor \citep{streamspray_original}. In \S~\ref{sec:discussion}, we discuss how to incorporate additional physics like multiple subhalos \citep{stream_gap}, a more physical tidal disruption that reproduces features like the stream cocoon \citep{Carlberg2018,Malhan2019,mass,Qian2022} and a time-dependent background potential \citep{Buist2015,Koppelman2021,Brooks2024}.

\begin{figure*}
  \centering
  \includegraphics[width=1\linewidth]{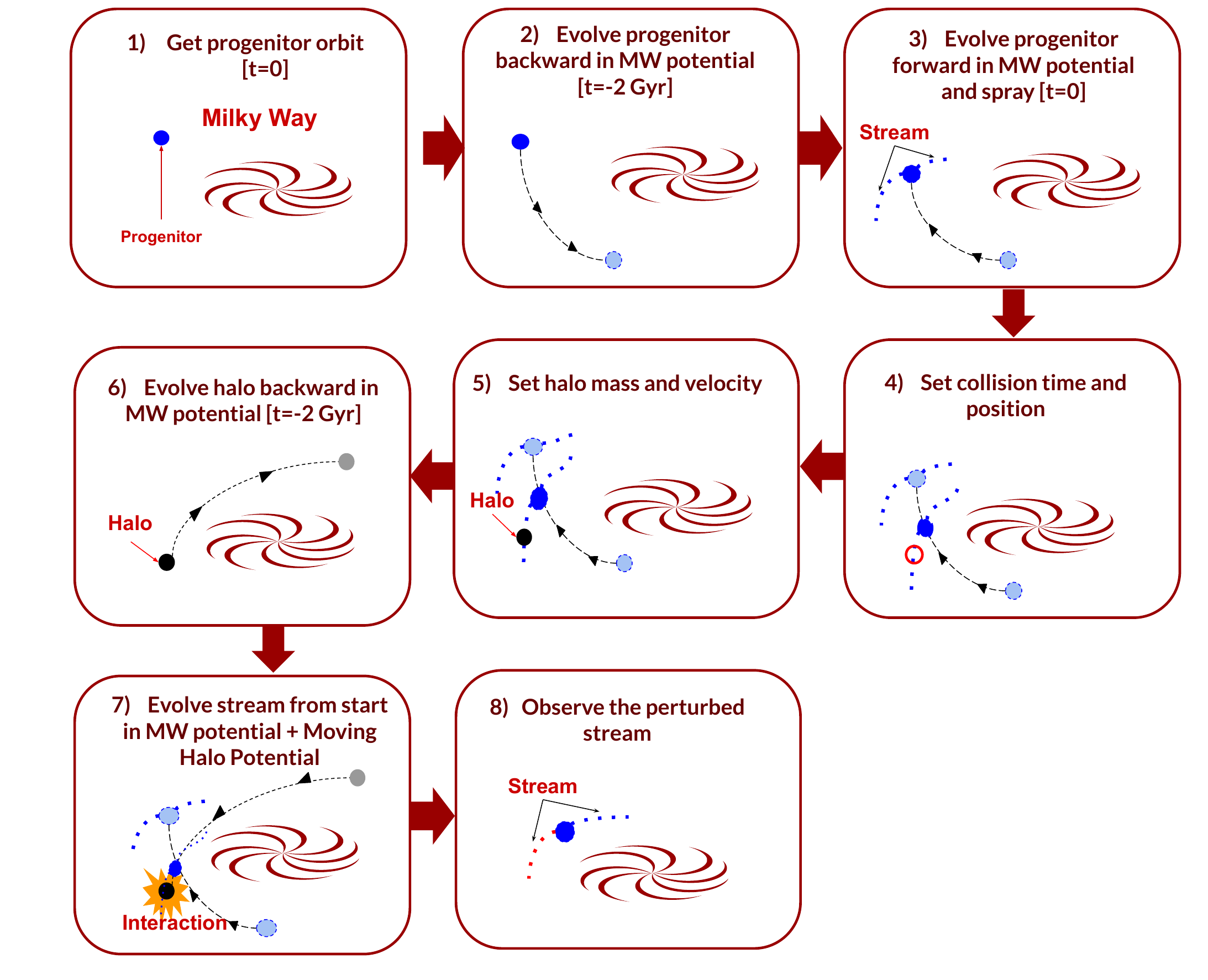}
  \caption{The steps in the stream simulation procedure (see main text for details).
  \label{fig:SIMULATION_STEPS}}
\end{figure*}

In this work, we set out to prove in principle whether the GCNN data compression and SBI inference method described above can reliably estimate the posterior distribution of the mass and velocity of a subhalo that interacted with a stellar stream. We describe the simulations, data compression/encoder and SBI in \S~\ref{sec:method}. In \S~\ref{sec:results}, we present the results of a comparison of our new approach to the current state-of-the-art 1D angular power spectrum analysis. In \S~\ref{sec:discussion}, we discuss our results and then conclude in \S~\ref{sec:concs}.

\section{Methods}\label{sec:method}

We describe the simulation procedure in \S~\ref{sec:sim} and the training set in \S~\ref{sec:datagen}. We describe the 1D angular power spectrum in \S~\ref{sec:PS} and the GCNN model in \S~\ref{sec:GCNN}. After compressing the simulated data, we present in \S~\ref{sec:sbi} the SBI method that we use to estimate the posterior distribution of the simulation parameters. In \S~\ref{sec:metrics}, we explain the performance metrics that we will use in \S~\ref{sec:results}.

\subsection{\textsc{streamspraydf} simulations}
\label{sec:sim}
In this proof-of-principle work, we simulate a {\gdone}-like stream interacting with a single subhalo in a static MW potential. We discuss more sophisticated simulations that we will use in future work in \S~\ref{sec:discussion_sims}. We use the \textsc{streamspraydf} algorithm \citep{streamspray_original} implemented in \textsc{galpy} \citep{galpy, Qian2022} to model {\gdone} by randomly ejecting stars from a progenitor according to the tidal disruption model of \cite{streamspray_original}. We integrate the orbit of each star within a fixed background gravitational potential, with and without the moving gravitational potential of a subhalo. For each training simulation described in \S~\ref{sec:datagen}, we vary the random realization of stars along the {\gdone} stream orbit and systematically vary the mass and velocity of the perturbing subhalo in order to span the prior parameter volume. In detail, for each simulation, we follow these steps (see also Fig.~\ref{fig:SIMULATION_STEPS}):

\begin{figure*}
    \centering

    \includegraphics[width=0.9\textwidth]{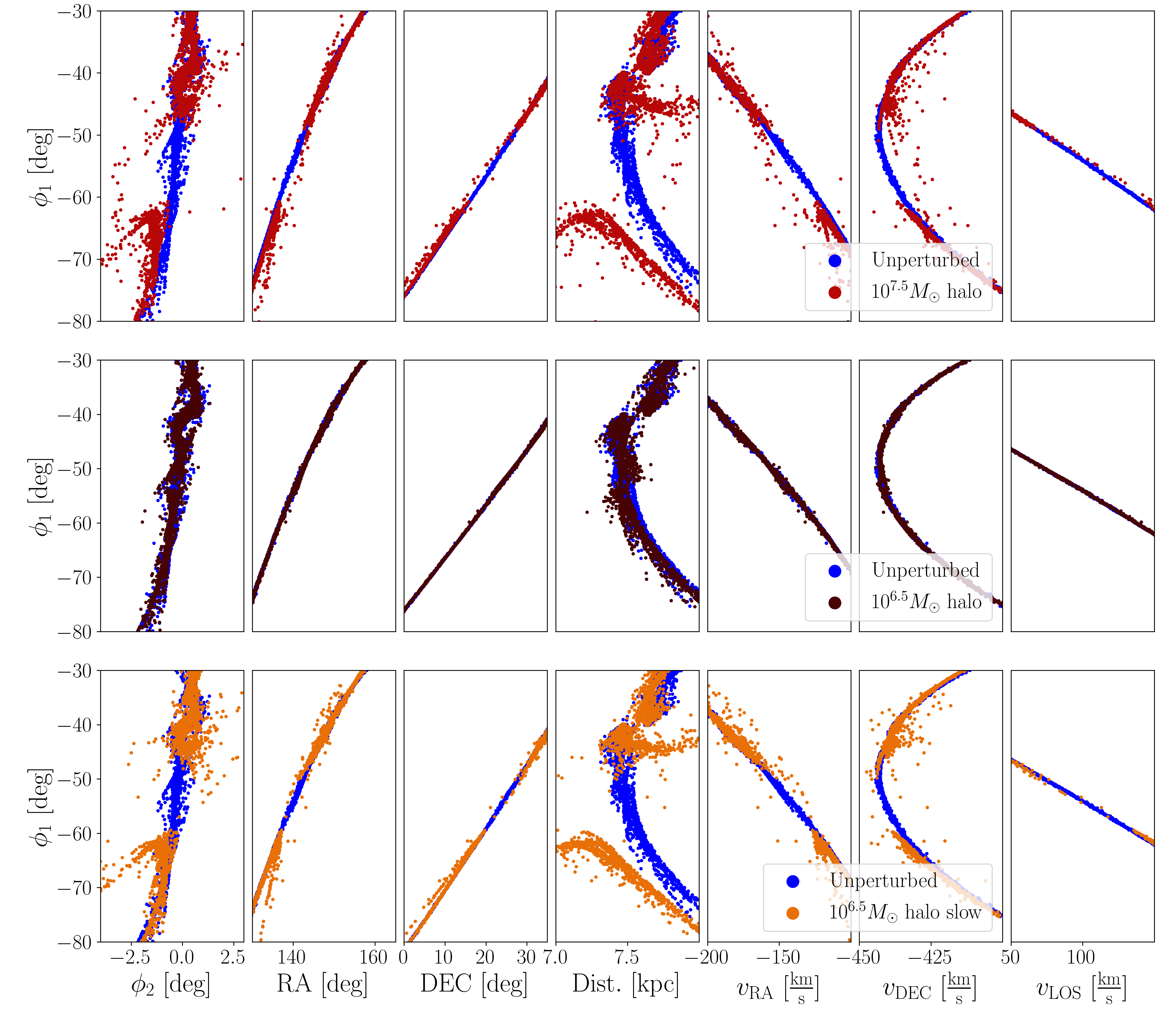}
    \includegraphics[width=0.85\textwidth]{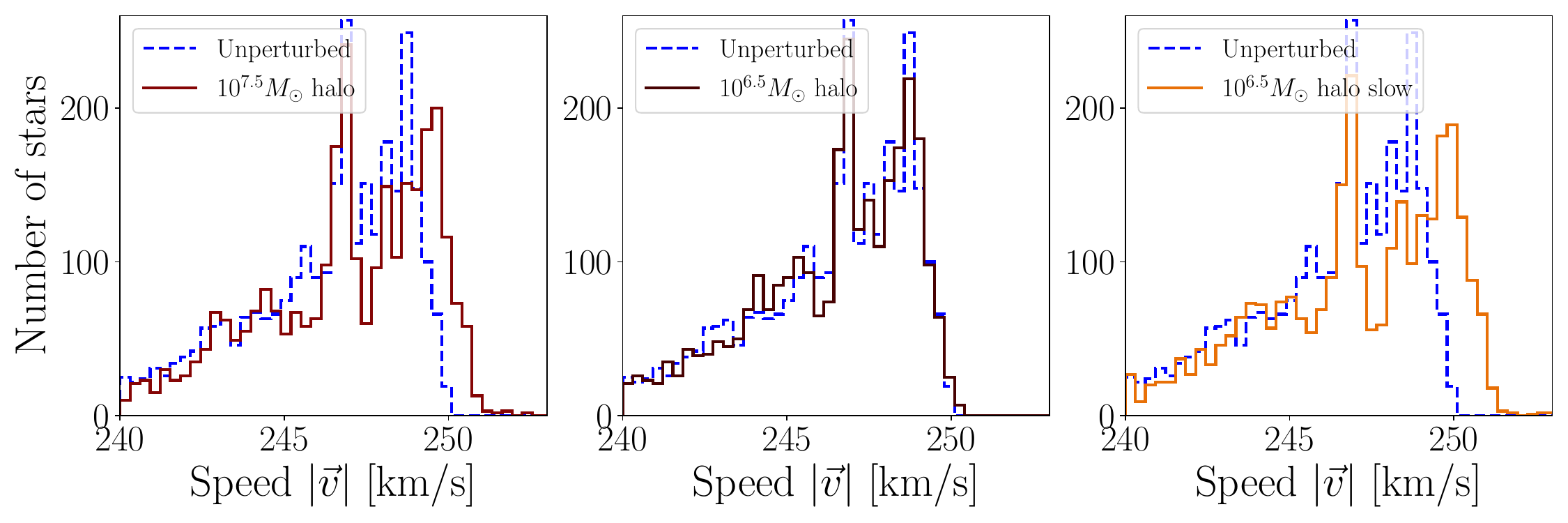}

    \caption{\textit{In the top three rows,} the phase space of the {\gdone} stellar stream simulations today. We compare the stream modeled without a subhalo (\textit{blue}) to the perturbed stream. In the top panel, the perturber is a subhalo of mass \(10^{7.5} M_\odot\) with a velocity of \(220\,\mathrm{km}/\mathrm{s}\) relative to the stream; in the middle, mass \(= 10^{6.5} M_\odot\) with the same velocity; in the bottom, mass \(= 10^{6.5} M_\odot\) with relative velocity \(= 22\,\mathrm{km}/\mathrm{s}\). \textit{From left to right}, we project the streams into the sky coordinates \(\phi_1\), \(\phi_2\), right ascension (RA), declination (DEC), heliocentric distance ({Dist.}), transverse velocity arising from the RA \(v_\mathrm{RA}\) and DEC \(v_\mathrm{DEC}\) components of the proper motion and the line-of-sight velocity \(v_\mathrm{LOS}\). To decorrelate the position and velocity vectors, we use the transverse velocity defined as the product of the proper motion and the heliocentric distance. \textit{In the bottom row,} the distribution of the speeds of stars in the same simulations as above.}
    \label{fig:stream_init}
\end{figure*}

\begin{enumerate}
    \item We set the state of the {\gdone} progenitor (mass, radius, position, velocity as determined in \cite{gd-1,Koposov2010,mass}) today at time $t = 0$. The progenitor mass is set to a constant $10^4 M_\odot$ and scaled radius of $0.01$ kpc.
    \item We integrate the progenitor orbit backwards from $t = 0$ in a static MW potential with no substructure until \(t=t_\mathrm{i}\) in the past. We choose $t_\mathrm{i} = -2\,\mathrm{Gyr}$ in order to match the observed length of the {\gdone} stream \citep{mass}. For the MW potential, we use the \textsc{MWPotential2014} setting in \textsc{galpy}, which combines a Navarro-Frenk-White \citep[NFW,][]{nfw} halo potential, a Miyamoto-Nagai disc potential \citep{Miyamoto} and a power-law density spherical bulge potential with an exponential cut-off \citep{powerlaw}.
    \item Using the \textsc{streamspraydf} algorithm \citep{streamspray_original}, we evolve the progenitor forward in time, ejecting stars during integration to obtain an evolved stream without a subhalo interaction. We eject $ n_\mathrm{stars} = 3000$ stars, with each stream arm containing 1500 stars, in order to have a converged estimate of the angular power spectrum.\footnote{We assess convergence as when statistical fluctuations in the power spectrum as the number of stars increases drops below 10\%.}
    \item We select a target star with which the subhalo will most closely interact. We choose the time of interaction to be -200 Myr.
    \item We set the state of the subhalo (mass, radius, position, velocity relative to the stream, impact parameter, angle of approach to the stream) at the time of interaction. We vary the mass and relative velocity in the training set (\S~\ref{sec:datagen}). For the subhalo potential, we use a Hernquist potential \citep{Hernquist1990} \textcolor{black}{with a scale radius of 10 pc}. \textcolor{black}{We define the impact parameter as the closest approach of the subhalo to the stream and we set it to zero.}
    \item We integrate the subhalo orbit backwards from the time of interaction in the MW potential until \(t_\mathrm{i}\) in the past.
    \item We evolve the progenitor forward in time again from \(t_\mathrm{i}\) to today, ejecting stars to form the {\gdone} stream, now with the background MW potential and the moving subhalo potential to model the subhalo interaction.
    \item The final result is the 6D phase space today of the stars in a {\gdone}-like stream.
\end{enumerate}

\begin{figure*}
  \centering
  \includegraphics[width=0.95\linewidth]{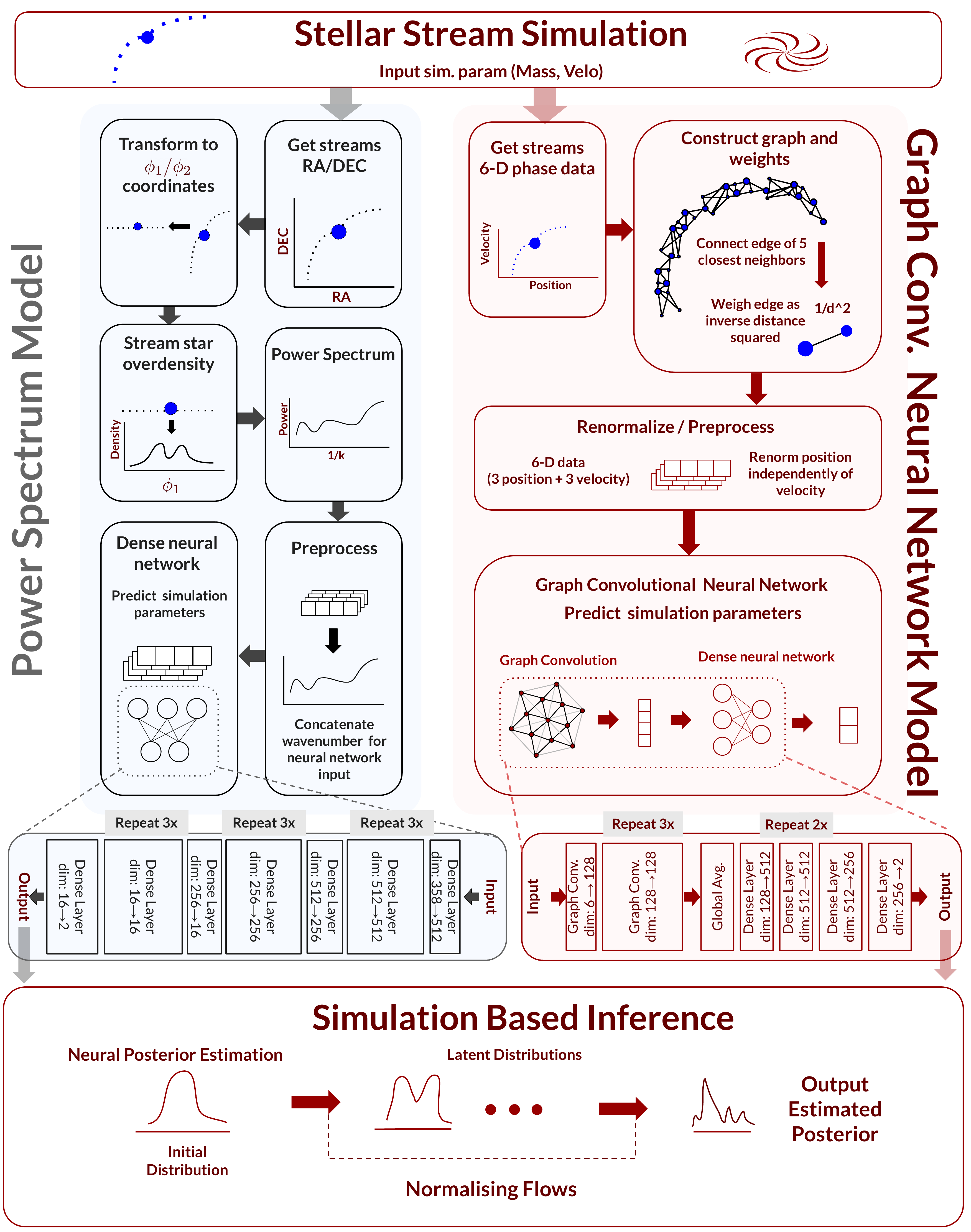}
  \caption{The steps in the encoder (\S~\ref{sec:encoder}) and simulation-based inference (\S~\ref{sec:sbi}) parts of the pipeline. We consider two possible encoders: (\textit{left}) the 1D angular power spectrum compressed by a dense neural network (\S~\ref{sec:PS}) and (\textit{right}) a graph convolutional neural network (GCNN) compressing directly from the stream phase space. The GCNN layer structure is specifically shown for the baseline GCNN model variant (Table \ref{tab:models}).
  }  
  \label{fig:model_design}
\end{figure*}

In Fig.~\ref{fig:stream_init}, we visualize examples of the simulated {\gdone} streams as they appear today with and without subhalo perturbations. We project the streams into right ascension (RA), declination (DEC) and line-of-sight position and velocity coordinates, as well as $\phi_1$ and $\phi_2$ coordinates which are transformations of RA and DEC using the rotation matrix defined in \citet{Koposov2010}. $\phi_1$ and $\phi_2$ are defined such that the stream has maximum extent along \(\phi_1\). A perturbing subhalo is qualitatively able to produce observed stream features like gaps, spurs and blobs \citep{stream}. The strength of the stream perturbation decreases as the subhalo decreases in mass to the extent that a gap may no longer form, but rather there is a reduction in the number density of stars near the point of interaction. This qualitative change in behavior occurs around \(10^7 M_\odot\) consistent with the previous simulations of \citet{stream}. However, there is a visual degeneracy between the effect of a slower, less massive subhalo and a faster, more massive subhalo. This effect arises because a slower subhalo spends more time closer to the stream and thus exerts a greater impulse on the stars. Thus, in order to break this degeneracy between subhalo mass and velocity, it will be crucial to extract as much information as possible from the full stream phase space, in particular the velocity distribution as shown in the bottom row of Fig.~\ref{fig:stream_init}. As expected, a subhalo perturbation produces a tail of high-speed stars in the stream. An animation of an example stream simulation can be found at \url{https://www.youtube.com/watch?v=d2nWvScdbJQ}.

\subsubsection{Generation of training simulations}
\label{sec:datagen}

For the machine learning (ML) in \S~\ref{sec:encoder} and \S~\ref{sec:sbi}, we generate two training sets, respectively with 40,000 and 80,000 simulations, in order to test how performance scales with the size of the ML model. Each training set is reused for the encoder and SBI models. In each set, we vary subhalo masses \(m\) distributed evenly in \textcolor{black}{the logarithm of mass for} the range $5 \leq \log{\frac{m}{M_\odot}} \leq 10$. We include masses down to the forecast observational limit of the \textit{Rubin} LSST survey \citep{lsstdm_review}. The upper prior bound excludes catastrophic interactions with massive satellites that would destroy the stream. We simultaneously vary subhalo velocities relative to the stream \(v\) \textcolor{black}{uniformly and linearly in the velocity for} the range $22 \leq \frac{v}{\mathrm{km}/\mathrm{s}} \leq 1100$. The upper prior bound arises as anything faster would leave no appreciable effect on the stream, while the lower bound approaches the limit where the subhalo is moving with the stream. In each simulation, we vary the random realization of stars that are ejected from the progenitor, but always with a progenitor orbit that matches {\gdone} observations. We sample the prior volume with a Latin hypercube  \citep{latincube}. The creation of the 40,000 simulation set takes approximately one week of computation on a 128 core machine with 1TB of RAM. The numerical integration in \textsc{galpy} uses a fourth-order Runge–Kutta method implemented in the \textsc{C} language with temporal discretization into 2000 intervals.

\begin{table*}[htbp!]
\centering
\begin{tabular}{|c|ccccc|}
\hline
\textbf{Model} & \textbf{Encoder} & \textbf{Coordinates}  & \textbf{\# sim.} & \textbf{\# stars} &\textbf{\# model pars.}  \\
\hline
PS & Power spectrum & $\phi_1$ &$4\times 10^4$ & 3000 & $7\times 10^5$\\
\hline
GCNN & Graph CNN & Galactocentric spherical 6-D &  $4\times 10^4$ & 3000 & $7\times 10^5$\\
\hline
GCNN-L  & Graph CNN & Galactocentric spherical 6-D  & $8\times 10^4$  & 3000 & $1.4\times 10^6$\\
\hline
GCNN helio (6-D)& Graph CNN & [RA, DEC, dist., $v_{\mathrm{RA}}$, $v_{\mathrm{DEC}}$, $v_{\mathrm{LOS}}$] &  $8\times 10^4$  & 3000 & $1.4\times 10^6$\\
\hline
GCNN helio (3-D)& Graph CNN & [RA, DEC, $v_\mathrm{LOS}$] &  $8\times 10^4$ & 3000 & $1.4\times 10^6$\\
\hline
 GCNN (6-D) 300 stars& Graph CNN &  [RA, DEC, dist., $v_{\mathrm{RA}}$, $v_{\mathrm{DEC}}$, $v_{\mathrm{LOS}}$] & $8\times 10^4$ & 300 & $1.4\times 10^6$\\
\hline
\end{tabular}
\caption{The encoder model variants that we consider. \textit{From left to right}, we give the model name, the type of encoder, the input coordinate system, the number of training simulations, the number of input stars, the number of encoder neural network hyperparameters that are trained.}
\label{tab:models}
\end{table*}

\begin{figure}
  \centering
  \includegraphics[width=1\linewidth]{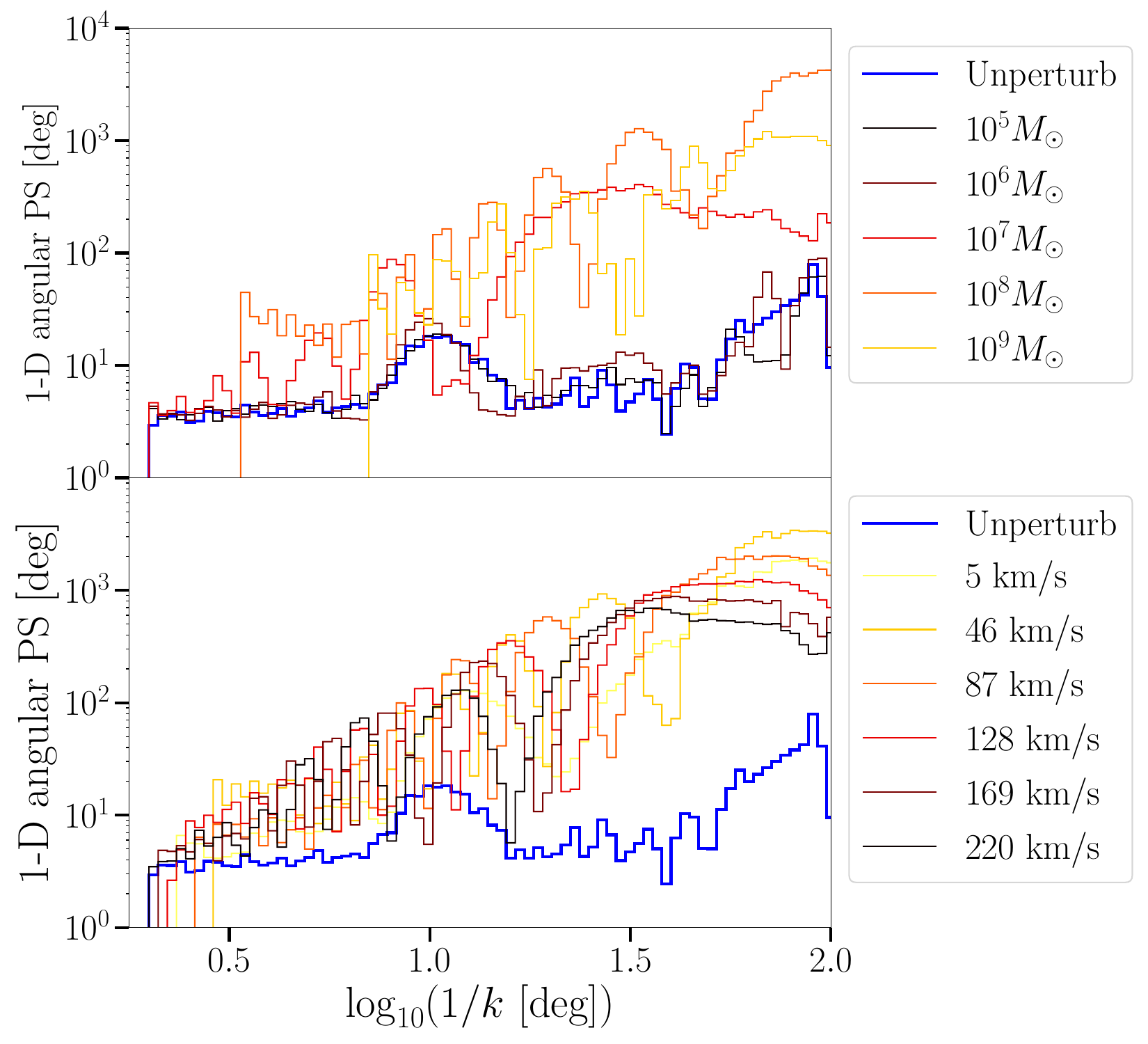}
  \caption{The 1D angular power spectrum \(P(k_{\phi_1})\) of the {\gdone} star number overdensity today as a function of the inverse of the angular wavenumber \(k_{\phi_1}\), for an unperturbed stream (\textit{blue}) and streams with different perturbers. \textit{In the top panel}, we vary the mass of the perturbing subhalo with velocity fixed to 110 km/s; \textit{in the bottom panel}, we vary the velocity of the subhalo with mass fixed to $10^{7.5} M_\odot$. Here, we show the power spectrum averaged over 100 random stream realizations in order to mitigate sample variance. A more massive and/or a slower subhalo increases the power on a scale of tens of degrees. A characteristic sinc pattern arises in the Fourier space from the tophat-like gap that forms in the stream density.}
  \label{fig:ps_mass_velo}
\end{figure}

\subsection{Encoder}
\label{sec:encoder}

The encoder is the layer of computation within our pipeline that compresses the stream phase space into a set of informative summary statistics. In this work, we choose to compress maximally to estimators of the simulation parameters, i.e., subhalo mass and velocity. We compare two encoders: the current state-of-the-art in data analysis, the 1D angular power spectrum of the star density (then compressed using a neural network to parameter estimators, PS, \S~\ref{sec:PS}); and a graph convolutional neural network in order to compress directly from the phase space (GCNN, \S~\ref{sec:GCNN}). For the GCNN encoder, we then introduce four model variants: (i) a larger GCNN with twice the number of training simulations (GCNN-L); (ii) a change of the input coordinate system from galactocentric to heliocentric (GCNN helio (6-D)); (iii) the effect of a reduced phase space (GCNN helio (3-D), only RA, DEC, \(v_\mathrm{LOS}\)); and (iv) a reduction in the number of stars from 3000 to 300 (GCNN (6D) 300 stars). We summarize these variants in Table \ref{tab:models}. We show in Fig.~\ref{fig:model_design} a schematic summary of the full pipeline from input simulations (\S~\ref{sec:sim}) through the two different encoders that we consider to the final simulation-based inference step (\S~\ref{sec:sbi}).

\subsubsection{1D angular power spectrum}
\label{sec:PS}
The 1D angular power spectrum [PS; \(P(k_{\phi_1})\)] of the star number overdensity \(\delta(\phi_1)\) quantifies the variance in the number density of stars as a function of angular wavenumber \(k_{\phi_1}\) along the extent of the stream:
\begin{equation}
    P(k_{\phi_1}) = \left|\hat{\delta}(k_{\phi_1})\right|^2 = \left|\int \delta(\phi_1) e^{-2\pi ik_{\phi_1} \phi_1}\mathrm{d}\phi_1 \right|^2.
    \label{eq:power}
\end{equation}
Here, \(\hat{\delta}(k_{\phi_1})\) is the Fourier transform of \(\delta(\phi_1)\) and $k_{\phi_1}$ is the variable conjugate to $\phi_1$. \(\delta(\phi_1) = \frac{n(\phi_1) - \overline{n}(\phi_1)}{\overline{n}(\phi_1)}\), where \(n(\phi_1)\) is the number density of stars and \(\overline{n}(\phi_1)\) is the average number density in the stream. The PS is sensitive only to Gaussian-distributed star perturbations in the \(\phi_1\) coordinate.

Figure \ref{fig:ps_mass_velo} demonstrates the sensitivity of the power spectrum to stream perturbers of different mass and velocity. We show an increase in power on larger angular scales as the mass of the perturber increases. This trend arises as the perturbation scatters the spatial distribution of stars, in particular on the angular scale of the gap that forms in the stream. Indeed, we see a characteristic sinc pattern in the power spectrum, which is the Fourier transform of the tophat-like gap that forms. We also show an increase in power on larger angular scales as the velocity of the perturber decreases. The less the velocity of the perturber, the more time the subhalo is close to the stream. This effect increases the impulse of the subhalo on the stream and scatters more stars. The same sinc pattern is generated, where a slower subhalo causes a larger gap leading to a more strongly oscillating pattern in Fourier space. There is thus a degeneracy in the power spectrum between increasing the perturber mass and decreasing its velocity.

In order to infer simulation parameters from the power spectrum, we must construct a likelihood function. Following, e.g., \citet{Banik2021_obs}, we do not assume a functional form for the likelihood\footnote{\citet{Banik2021_obs} use a form of SBI called approximate Bayesian computation.} and instead use simulation-based inference (\S~\ref{sec:sbi}), which also allows a more direct comparison to the GCNN approach (\S~\ref{sec:GCNN}). In order to use SBI, we must reduce the dimensionality of the simulated data further as the density estimation task in SBI scales with the sum of the dimensions of the data and parameters. We do this compression using a dense neural network to compress the power spectrum to two numbers, which are estimators of the subhalo mass and velocity.

We proceed by normalizing \(\log P(k_{\phi_1})\) to the unit interval as input to the neural network. In order also to feed the neural network information on the wavenumber dependence of the power spectrum, we concatenate to the power spectrum vector, a vector of inverse wavenumbers \(\frac{1}{k_{\phi_1}}\) at the centers of the bins that we use, also normalized to the unit interval. We use 179 bins distributed evenly in \(\log \frac{1}{k_{\phi_1}}\), thus meaning that the first layer of the network has 358 elements.

The dense neural network applies to the input vector a series of composed parameterized functions (layers). In succession, each \(n\)\textsuperscript{th} layer $\textbf{l}_{n}$ first linearly transforms the previous layer output \(\textbf{l}_{n-1}\) by matrix multiplying an optimizable weight matrix $\textbf{W}^{(n)}$ and then adding an optimizable bias vector \(\textbf{b}^{(n)}\) before applying a non-linear activation function \(\alpha\) \citep{nn_func_approx}:
\begin{equation}
    \textbf{l}_n = \alpha\left(\textbf{W}^{(n)}\textbf{l}_{n-1}+ \textbf{b}^{(n)}\right).\label{eq:nn_layer}
\end{equation}
The structure of these layers, i.e., the number of elements in each, is shown in Fig.~\ref{fig:model_design}, while the details are given in Appendix \ref{sec:hyperparam}. We use Leaky ReLU as the activation function. The weights and biases are optimized by minimizing a loss function \(\mathcal{L}(\Vec{x}, \Vec{\theta}; p_\gamma)\) of the final layer. We use the mean absolute error (MAE) between the output of the network \(p_\gamma(\Vec{x})\) and the true subhalo parameters \(\Vec{\theta} = [m, v]\) (normalized to the prior volume defined in \S~\ref{sec:datagen}) for a set of training simulations:
\begin{equation}
    \mathcal{L}(\Vec{x}, \Vec{\theta}; p_\gamma) = \frac{1}{2}\sum_{i=1}^2\left|p_\gamma^{(i)}(\Vec{x}) - \theta_i\right|,
    \label{eq:mae}
\end{equation}
where \(\Vec{x} = \left[P(k_{\phi_1}), \frac{1}{k_{\phi_1}}\right]\) is the input vector to the network, \(p_\gamma^{(i)}\) represents the neural network transformation of the input and \(i\) indexes over the two model parameters (mass and velocity).

We use backpropagation to train the network \citep{backprop}. Backpropagation is an efficient means of computing the gradients of the loss with respect to the network's hyperparameters. This method is useful for the gradient-based optimization algorithm, adaptive moment estimation \citep[\textsc{ADAM},][]{adam}, that we use. In building the neural network, we optimize its configuration by performing a neural architecture search. We do this using Bayesian optimization (see Appendix \ref{sec:hyperparam}). We split the input simulation set (see \S~\ref{sec:datagen}), with 80\% for training and 20\% for validation; the test set has $9000$ simulations drawn as a Latin hypercube across the prior area. We use the \textsc{ADAM} optimizer with a constant learning rate of $10^{-3}$ and a batch size of $32$. We train the network until saturation (early stopping). We give additional training details in Appendix \ref{sec:training}. 

\begin{figure}
  \centering
  \includegraphics[width=1\linewidth, trim={0cm, 0cm, 0cm, 3.5cm}]{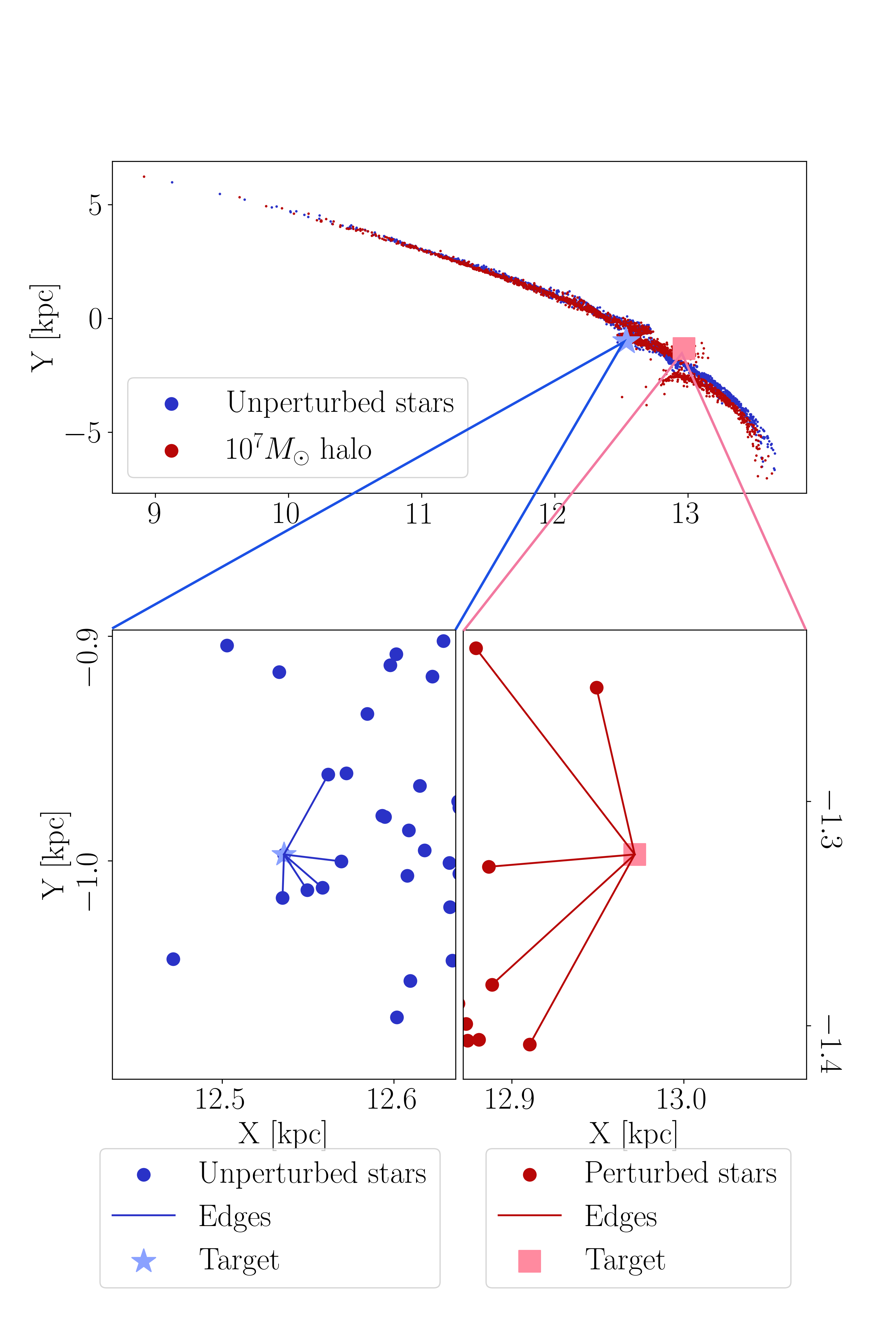}
  \caption{The \textit{top panel} projects {\gdone} into galactocentric X and Y coordinates, with \textit{blue} points showing the positions of stars in an unperturbed realization and \textit{black} points showing stars in a realization with a perturbing subhalo of mass \(10^7 M_\odot\) (same random seed in each realization). In the \textit{bottom panels}, we show zoom-ins on the same target star in each realization (unperturbed on the \textit{left} and perturbed on the \textit{right}). In the zoom-ins, we show the graph that we construct for the target star, connecting the star with edges to the five nearest neighbors. We illustrate how the graph is sensitive to the geometry of the stream that the perturbation induces.}
  \label{fig:graph_struct}
\end{figure}

\subsubsection{Graph convolutional neural network}
\label{sec:GCNN}
As a less lossy alternative to the power spectrum (\S~\ref{sec:PS}), we also compress the stream to subhalo parameter estimators using a neural network directly applied to the phase space, i.e., without the power spectrum ``pre-compression''. Rather than use a dense neural network as described in \S~\ref{sec:PS}, we use a graph convolutional neural network, at least for the first part of the pipeline. First, we construct a graph from the stream phase space.

In Fig.~\ref{fig:graph_struct}, we show an example of the graphs that we construct (specifically, part of the total stream graph for a single target star). To form a graph, we connect each star in the stream, as a node, to its $k$-nearest neighbors using an edge, where we set $k=5$. At each node, we list the features of each star as its 6-D phase space position (or a subset of the full 6-D information depending on the model variant, see Table \ref{tab:models}), i.e., its position and velocity vector in the chosen coordinate system. Each edge \(e_{ij}\) is weighted by the square of the inverse distance from star \(i\) to neighboring star \(j\). In Fig.~\ref{fig:graph_struct}, we highlight a single star and its connections and see that graphs capture local geometric structures that are highly sensitive to subhalo perturbations. Edges connecting perturbed stars (\textit{bottom right panel}) are longer compared to those connecting unperturbed stars (\textit{bottom left panel}). By connecting only the five closest neighbors, the graph is kept sparse which generates deeper networks and avoids oversmoothing \citep{smoothing}. We observe that smaller values of $k$ lead to a lack of connectivity, which degrades training performance. Therefore, we choose $k=5$ as a fixed hyperparameter.

We proceed by normalizing the input features (i.e., the phase space vector) at each graph node by the standard deviation of each feature across the training set. Having constructed a graph with normalized features from the stream phase space, we then pass the graph as input to a GCNN. We first apply a series of graph convolutional layers. Each convolutional layer applies three operations: a message passing (where the features at each node are replaced by an aggregation of the features at connected nodes), a linear affine transformation (the application of a weight matrix and bias term) and a non-linear activation (these last two operations are identical to the neural network in Eq.~\eqref{eq:nn_layer}).

In other words, each \(n\)\textsuperscript{th} layer is now a matrix \(\textbf{l}_n\) consisting of features (along rows) for each node (stars along columns). First, the previous layer \(\textbf{l}_{n-1}\) is matrix multiplied by a symmetric adjacency matrix \(\textbf{A}\) that encodes the message passing between connected nodes (i.e., the graph convolution). Then, as before, this convolved layer is matrix multiplied by an optimizable weight matrix \(\textbf{W}^{(n)}\) and then added to a bias term, which is now a bias matrix \(\textbf{b}^{(n)}\), i.e., a set of biases (along rows) for each node (stars along columns). Finally, we apply a non-linear activation function \(\sigma\):
\begin{equation}
    \textbf{l}_n = \sigma\left(\textbf{A}\textbf{l}_{n-1}\textbf{W}^{(n)} + \textbf{b}^{(n)}\right).
    \label{eq:gcnn_layer}
\end{equation}

The adjacency matrix \(\textbf{A}\) has elements:
\begin{equation}
    A_{ij} = 
    \begin{cases}
    e_{ij}, \quad \text{if \(i\)\textsuperscript{th} node connected to \(j\)\textsuperscript{th} node,}\\
    0, \quad \text{else.}
    \end{cases}
    \label{eq:adj}
\end{equation}
In this way, the adjacency matrix performs the graph convolution by, at each node, replacing each feature by a weighted sum of the equivalent feature at connected nodes. This is analogous to an image convolution, as applied in a standard convolutional neural network, where each pixel in each layer is an aggregation of nearby pixels weighted by a kernel. Here, the image kernel is replaced by the adjacency matrix.

The structure of the GCNN layers, i.e., the number of features at each node, is shown in Fig.~\ref{fig:model_design}, while the details are given in Appendix \ref{sec:hyperparam}. Our choice of edge weighting has the property that $e_{ij}= e_{ji}$ which enforces equivariance symmetry. This bi-directionality means that pairwise force interactions are captured in the compression network.

After the graph convolutional layers, we apply a layer that, for each of the final 128 features, averages the feature across all nodes (stars).\footnote{We visualize the final graph convolutional layer in Appendix \ref{sec:visualise}.} We then pass this single vector of features through a series of standard layers [Eq.~\eqref{eq:nn_layer}], analogous to the dense neural network described in \S~\ref{sec:PS}. The final layer is again estimators of the two subhalo parameters. The structure of these final layers is also given in Fig.~\ref{fig:model_design}, while the details are again given in Appendix \ref{sec:hyperparam}. We otherwise follow the same steps as in \S~\ref{sec:PS}, i.e., we use the same loss function, optimizer, backpropagation, early stopping and training-validation split. We again use Bayesian optimization to optimize the architecture configuration (see Appendix \ref{sec:hyperparam}, meaning that the activation function is instead ReLU); we give additional training details in Appendix \ref{sec:training}. The training of the baseline GCNN model takes 19 hours on an NVIDIA RTX 6000 Ada Generation graphics processing unit. The slight modifications to the structure of the layers necessary for the GCNN encoder model variants (Table \ref{tab:models}) are explained in \S~\ref{sec:scale_gcnn} and \ref{sec:coord_obs}.

\subsection{Simulation-based inference}
\label{sec:sbi}

Having compressed the stream phase space by the two alternative encoders presented in \S~\ref{sec:encoder}, we now proceed to estimating the posterior distribution \(\mathcal{P}(\Vec{\theta}|\Vec{\hat{\theta}})\) of the simulation parameters \(\Vec{\theta} = [m, v]\) given these compressed summary statistics \(\Vec{\hat{\theta}} = p_\gamma(\Vec{x})\). Rather than assuming a simple approximation for the posterior or likelihood (e.g., a Gaussian function), we instead use SBI to learn a parameterized form of the posterior from a training set of forward models (simulations).

We use neural posterior estimation, as implemented in the \texttt{sbi} package \citep{sbi-toolkit}, which, using neural networks, directly learns the posterior, rather than other approaches that learn the likelihood \citep{likelihood_est} or the ratio of likelihood to evidence \citep{ratio_est, ratio_est_2}. The parameterized distribution that we use to model the posterior is a normalizing flow \(\mathcal{P}_\phi(\Vec{\theta}|\Vec{\hat{\theta}})\) \citep{posterior_est_1, posterior_est_2, norm_flow, posterior_est_3}. Normalizing flows take a simple base distribution (in our case, a Gaussian) and apply a series of parameterized, invertible, volume-preserving (normalized) transformations to match a more complex target distribution. In our case, the target is the true posterior \(\mathcal{P}(\Vec{\theta}|\Vec{\hat{\theta}})\). Specifically, we implement a masked autoregressive flow \citep{MAF}. Here, we factor the posterior distribution into a series of conditional probabilities via the chain rule of probabilities (autoregressive). When training the flow, we mask variables to ensure the sequential variable ordering so we obey the chain rule.

In order to learn the parameters \(\phi\) of this flow model, we use a neural network with simulated estimators \(\Vec{\hat{\theta}}\) and true parameters \(\Vec{\theta}\) as input. We optimize this neural network by minimizing the divergence between the true and target distributions. This minimization is equivalent to minimizing a loss function \(\mathcal{L}(\Vec{\hat{\theta}}, \Vec{\theta})\) that is the negative log posterior for the training samples \(\{\Vec{\theta}_i,\Vec{\hat{\theta}}_i\}\):
\begin{equation}
    \mathcal{L}(\Vec{\hat{\theta}}, \Vec{\theta}) = \sum_i - \ln \mathcal{P}_\phi(\Vec{\theta}_i|\Vec{\hat{\theta}}_i),
    \label{eq:flow_loss}
\end{equation}
where, here, \(i\) indexes over the training simulations.

The training, validation and test simulation split follows the same proportions as for the encoder models. We optimize the SBI model using \textsc{ADAM} with a constant learning rate of $5 \times 10^{-4}$. Otherwise, we use the default \texttt{sbi} settings. Having learnt a model for the posterior distribution, we draw samples for a given mock data vector in order to generate the posterior intervals in \S~\ref{sec:results} \citep{reject}.

\subsection{Performance metrics} \label{sec:metrics}

In \S~\ref{sec:ps_gcnn_compare}, we compare the performance of the power spectrum (\S~\ref{sec:PS}) and GCNN (\S~\ref{sec:GCNN}) encoders by calculating the fractional mean absolute error on the test set:
\begin{equation}
   \mathrm{MAE}_{\mathrm{frac},i}  \equiv \frac{\left|\hat{\theta}_i -  \theta_i\right|}{\theta_i},
\label{eq:frac_err}
\end{equation}
where \(i\) indexes the two simulation parameters (subhalo mass and velocity) and \(\hat{\theta}_i\) is the parameter estimator given either the power spectrum or GCNN compression. The smaller this metric, the more informative is the compressed summary statistic \(\hat{\theta}_i\).

\begin{figure}
\begin{center}

\includegraphics[width=\columnwidth]{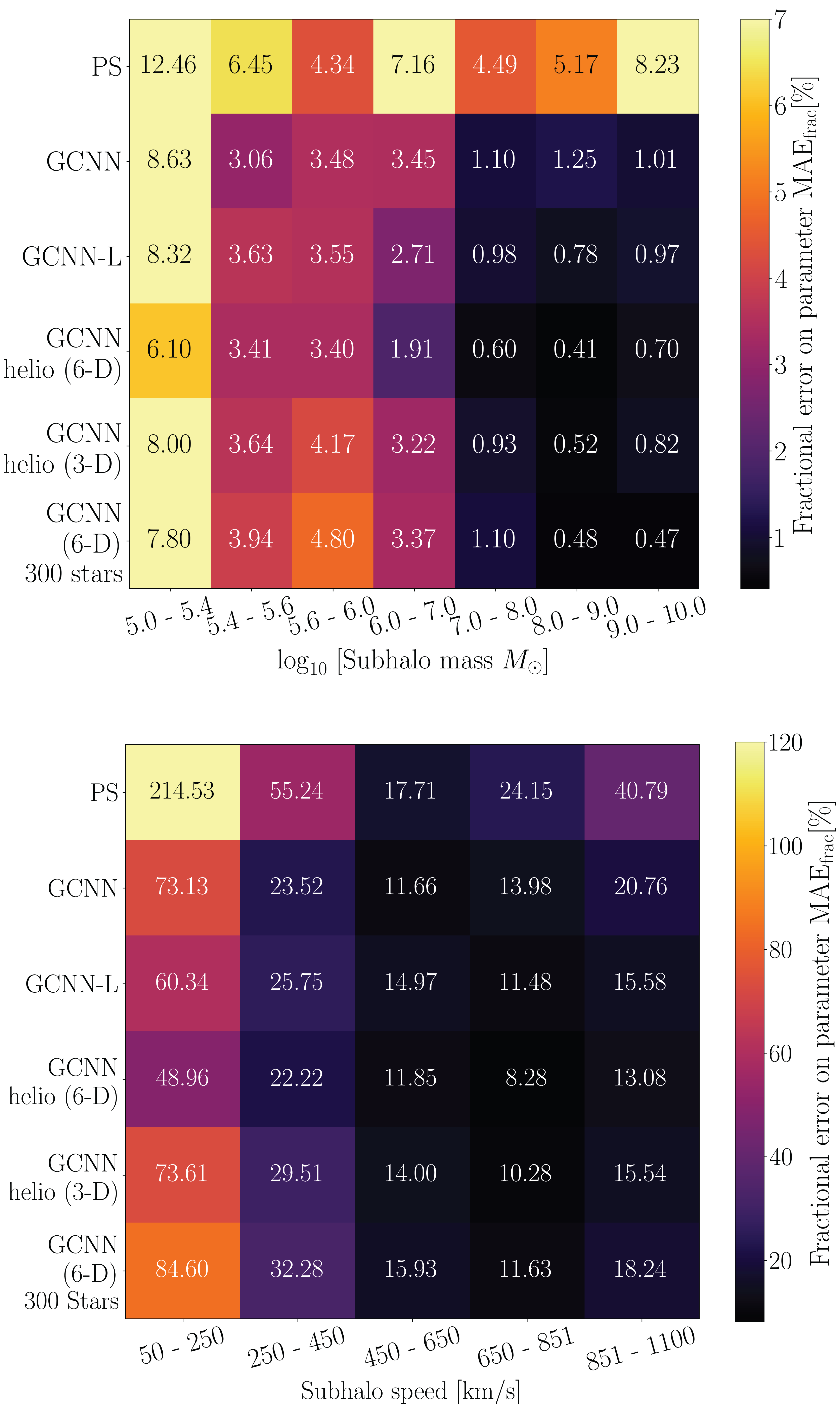}
\caption{\textit{In the top panel,} the fractional mean absolute error [as shown in Eq.~\eqref{eq:frac_err}] on the true subhalo mass for a test set of stream simulations. \textit{From left to right} in this panel, we show \(\mathrm{MAE}_\mathrm{frac}\) averaged over all test samples in the given mass bin\textcolor{black}{; each mass bin averages across all velocity bins}. \textit{From top to bottom} in this panel, we vary the encoder model variant (see Table \ref{tab:models}). \textit{In the bottom panel,} same as the top panel except computed for bins of subhalo speed\textcolor{black}{; each velocity bin averages across all mass bins}.}
    \label{fig:mae_grid}
    \end{center}
\end{figure}

\begin{figure*}
    \centering\includegraphics[width=\textwidth, trim={0cm, 0cm, 0cm, 1cm}]{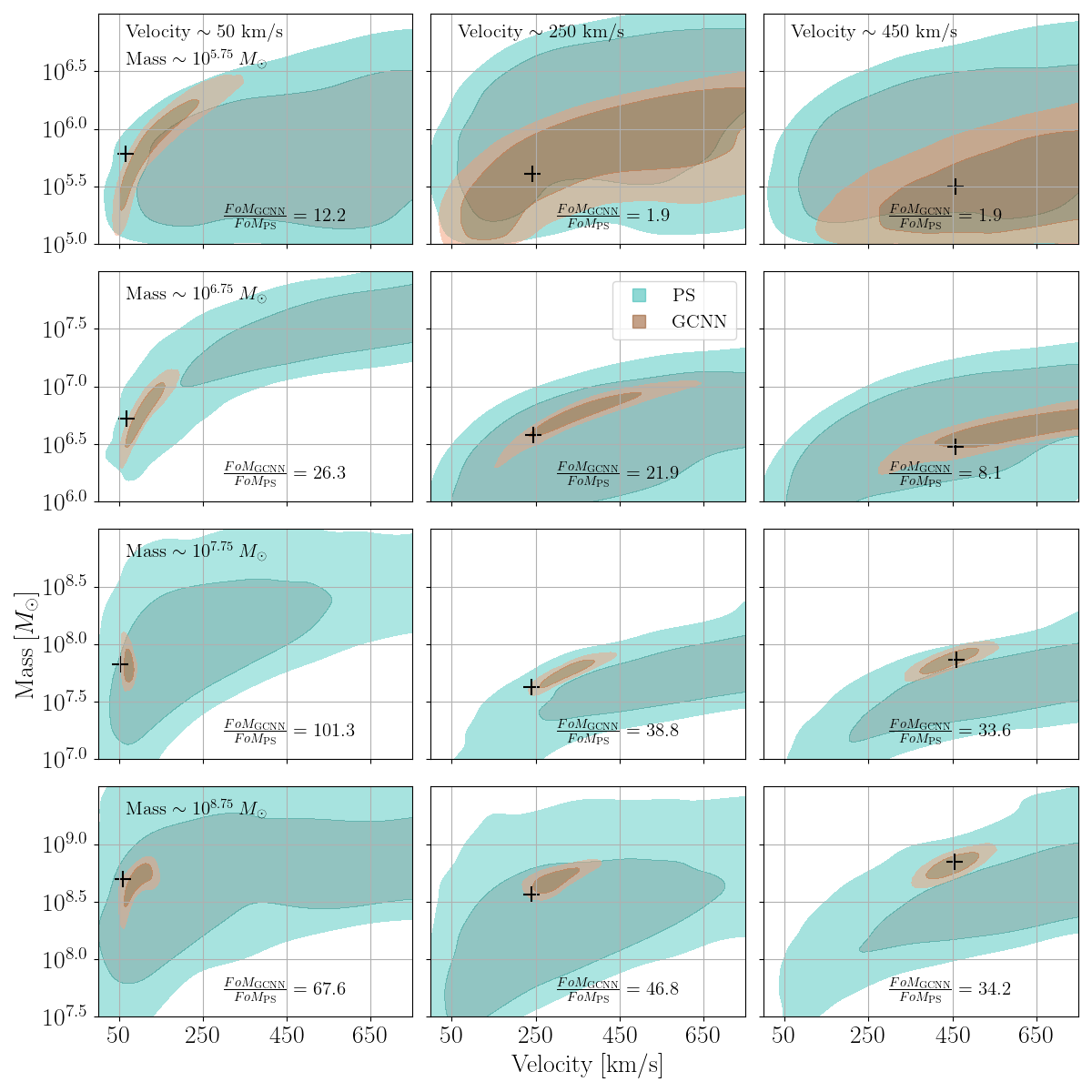}
        \caption{The posterior distribution as estimated by SBI (\S~\ref{sec:sbi}) given a test set of stream simulations with true parameter values indicated by the black crosses. The \textit{cyan} contours show posteriors given the power spectrum encoder; \textit{brown} contours show posteriors given the baseline GCNN encoder (see Table \ref{tab:models}). \textit{From top to bottom}, we vary the subhalo mass of the test simulation (see labels in left-hand column); \textit{from left to right}, we vary the subhalo velocity of the test simulation (see labels in top row). The darker and lighter contours respectively indicate the 68\% and 95\% credible intervals. \textit{In each panel}, we give the ratio of the figures of merit [Eq.~\eqref{eq:FoM}] given the two encoders. The figure of merit (the posterior area) is always improved (smaller) when using the GCNN encoder, with the greatest improvement at higher mass and/or lower velocity. There is some scatter in the figure of merit ratio and the position of the truth relative to the posterior owing to the noise in each test simulation. In \S~\ref{sec:SBC}, we confirm that all these posteriors are well calibrated.}
    \label{fig:post}
\end{figure*}

In \S~\ref{sec:SBI_calibration}, we compare the performance of the SBI (\S~\ref{sec:sbi}) given power spectrum and GCNN encoders, with regards to both the precision and accuracy of estimated posterior distributions. In order to quantify the improvement in the precision of the subhalo mass constraint, we calculate the ratio of the subhalo mass posterior standard deviation \(\sigma\) relative to the best-performing encoder \(\sigma_\mathrm{best}\), which is the GCNN helio (6-D) model variant (Table \ref{tab:models}):
\begin{equation}
\label{eq:sigma_ratio}
\mathrm{\sigma}_{\mathrm{frac}}  \equiv \frac{\sigma}{\sigma_{\mathrm{best}}}.
\end{equation}
In other words, we give this ratio such that larger values indicate that the marginalized posterior on the subhalo mass is broader (than the best case) and can conclude that the encoder has extracted less information from the stream. We also sometimes (Fig.~\ref{fig:post}) calculate the full figure of merit for the 2D posterior:
\begin{equation}
\label{eq:FoM}
FoM = \frac{1}{\sqrt{\mathrm{det}\,\textbf{C}}},
\end{equation}
where \textbf{C} is the posterior covariance between mass and velocity. The figure of merit quantifies the area of the peak of the posterior. By comparing this quantity, we assess the total information gain accounting for both subhalo mass and velocity.

We also assess the accuracy of the estimated posterior distribution, i.e., how correctly has the SBI procedure calculated the true posterior distribution given the input compressed data? First, we report the negative log probability of the true parameters averaged over the test set, i.e., the loss function [Eq.~\eqref{eq:flow_loss}] evaluated at the true parameters of the test set. The lower this metric, the higher probability is the true point (averaged over the test set), which is indicative of a better calibrated set of posteriors \citep{pmlr-v130-lueckmann21a}.

\begin{table*}
\begin{tabular}{|c|cccc|}
\hline
\textbf{Encoder model} & $\mathbf{\left(10^{5.75} - 10^6\right) M_\odot}$  & $\mathbf{\left(10^6 - 10^7\right) M_\odot}$ & $\mathbf{\left(10^7 - 10^8\right) M_\odot}$  & $\mathbf{\left(10^8 - 10^9\right) M_\odot}$ \\
\hline
PS & 1.70 $\pm$ 0.02 & 2.65 $\pm$ 0.25 & 6.73 $\pm$ 0.76  & 10.55 $\pm$ 1.00  \\
\hline
GCNN  & 1.15 $\pm$ 0.01 & 1.30 $\pm$ 0.01 &  1.67 $\pm$ 0.28 &  2.02 $\pm$   0.10 \\
\hline
GCNN-L  & 1.15 $\pm$ 0.05 & 1.23 $\pm$ 0.01 &1.47 $\pm$ 0.15 & 1.83 $\pm$ 0.29 \\
\hline
GCNN helio (6-D) & 1&1 & 1& 1 \\
\hline
GCNN helio (3-D) & 1.18 $\pm$ 0.03  &  1.37 $\pm$ 0.04 & 1.45 $\pm$  0.03  & 1.10 $\pm$ 0.07 \\
\hline
GCNN (6-D) 300 stars & 1.18 $\pm$ 0.06 & 1.48 $\pm$  0.14 &1.57 $\pm$  0.16  & 0.97 $\pm$ 0.11 \\
\hline
\end{tabular}
\caption{The ratio [as defined in Eq.~\eqref{eq:sigma_ratio}] of the inferred subhalo mass uncertainty (marginalized posterior standard deviation for a test set of stream simulations) for the encoder model variant given in the left hand column over the best performing case [GCNN helio (6-D)]. As discussed in \S~\ref{sec:SBC}, we report posteriors only for subhalo masses \(5.75 \leq \mathrm{log} \frac{m}{M_\odot} \leq 9\) and speeds \(27 \leq \frac{v}{\mathrm{km}/\mathrm{s}} \leq 450\), where the GCNN encoder distributions pass the simulation-based calibration test. The exception to this is the GCNN (6-D) 300 stars variant, where we report for \(27 \leq \frac{v}{\mathrm{km}/\mathrm{s}} \leq 350\). We give the mean and standard deviation of this ratio for test samples from the subhalo mass range given on the top row (and for speeds in the well-calibrated range given above). Subhalo mass constraints are always stronger given GCNN encoders than the power spectrum approach.}
\label{tab:sbi_test}
\end{table*}

Further, we perform a simulation-based calibration test \citep{sbi-cali,sbi-cali-2}. This test checks whether the peaks of posterior distributions generated by the SBI model are consistent with the true simulation parameters with the expected frequency; or, in other words, whether the credible intervals of the posterior have good coverage probabilities. We compute the posterior distribution given each test stream. We then draw samples from the posterior distribution and calculate a rank by counting how many samples fall below the true simulation parameter in terms of the value of the posterior probability. A well-calibrated set of posteriors will have a uniform distribution of true value ranks. If the rank distribution deviates from uniformity by having many ranks in the tails, this indicates that the posterior area is typically underestimated, i.e., over-confident posterior constraints or over-fitting. If the rank distribution has many ranks in the center, this indicates that the posterior area is typically overestimated, i.e., under-confident constraints or under-fitting.

We quantify the consistency of the measured rank distribution with a uniform distribution by the Kolmogorov-Smirnov (KS) test \citep{Kolmogorov, Smirnov, sbi-cali-2}. The KS test returns $p$ values on the null hypothesis that the measured ranks are drawn from a uniform distribution. We reject this null hypothesis if $p < 0.05$; we otherwise report that the posteriors are well calibrated. In practice, we find that posteriors near the edge of the prior can be poorly calibrated (owing to discontinuity in the uniform prior that we use) and so we report only a central area for each encoder model where \(p \geq 0.05\).

\section{Results}\label{sec:results}

\subsection{Encoder performance} \label{sec:ps_gcnn_compare}
Figure \ref{fig:mae_grid} evaluates the performance metric defined in Eq.~\eqref{eq:frac_err} (the fractional mean absolute error on the estimated subhalo mass and velocity) given the different encoder model variants defined in Table \ref{tab:models}. When averaged over all test samples, the power spectrum approach (\S~\ref{sec:PS}) has an absolute validation MAE [Eq.~\eqref{eq:mae}] of $0.157$. In comparison, the GCNN model (\S~\ref{sec:GCNN}) has an absolute validation MAE of $0.064$. The lower the MAE, the more informative is the data compression. This conclusion follows since the subhalo mass and velocity completely characterize each simulation, barring the incompressible noise of random star ejections from the stream progenitor given the tidal disruption model that we use (\S~\ref{sec:sim}).

Figure \ref{fig:mae_grid} breaks down this information gain by varying subhalo mass and velocity. In this section, we consider only the first two rows of each subpanel in Fig.~\ref{fig:mae_grid}, i.e., we compare the power spectrum approach with the baseline GCNN model. We consider the effect of increasing the size of the GCNN model in \S~\ref{sec:scale_gcnn} (GCNN-L) and the effects of coordinate system and incomplete observations in \S~\ref{sec:coord_obs} [GCNN helio (6-D), GCNN helio (3-D), GCNN (6-D) 300 stars].

The power spectrum model degrades in performance towards the edges of both the subhalo mass and velocity prior ranges. There are two physical effects that contribute to this behavior. First, when the mass is large and/or the velocity is low, the impulse of the subhalo on the stream is larger and so the stream is far more disrupted. Indeed, this disruption can occur to such an extent that the perturbed stars are no longer confined to variations in one angular coordinate \(\phi_1\), instead spreading out in all spatial dimensions. In this case, the 1D angular power spectrum becomes an increasingly poor compression of the stream phase space.

On the other hand, when the mass is small and/or the velocity is high, the impulse on the stream is smaller and so the stream is far less disrupted. In this case, indeed all encoder model variants perform increasingly poorly simply because there is less and less information in the stream about the weak interaction. A \(10^5 M_\odot\) subhalo is considered about the lower limit of what can be detected in a single stellar stream using upcoming observatories \citep[][although the details of this forecast depend on many other properties of the stream and its perturbers]{lsstdm_review}. Finally, the fractional MAE for the subhalo velocity is much larger than for the subhalo mass (again for all encoder model variants). This means that the stream phase space is more sensitive to variations in subhalo mass than velocity.

The baseline GCNN model performs better than the power spectrum in all twelve subhalo mass and velocity bins in Fig.~\ref{fig:mae_grid}. This improvement reflects that the GCNN produces more informative summary statistics across the prior area, largely driven by having more information as input, i.e., the full 6-D phase space rather than only 1-D for the power spectrum. The same physical effects from varying subhalo mass and velocity, as above, determine variations in the GCNN performance.

There are also two numerical effects from the encoder model training. First, performance tends to degrade at the lower ends of the prior ranges because the absolute MAE [Eq.~\eqref{eq:mae}] is smaller meaning that training and validation samples in this region contribute less to the total model loss function. This effect arises despite the normalization of training data described in \S~\ref{sec:encoder}. Second, performance also tends to degrade at the edges of the prior ranges because the number of nearby training simulations decreases, i.e., there is half the training density at the edge than at the center of the prior. The first numerical effect could be ameliorated by using a different loss function, e.g., the fractional MAE that we use as a performance metric in this section. The second numerical effect could be ameliorated by either varying the density of training samples across the prior range or extending the prior range so that the parameter space of physical interest always remains well sampled. Having demonstrated our primary objective, that the GCNN can consistently extract more information than the power spectrum, we defer to future work these fine tunings of the model training.

\subsection{Posterior distributions} \label{sec:SBI_calibration}
Figure \ref{fig:post} shows examples of posterior distributions given test stream data, comparing the use of the power spectrum and baseline GCNN encoders (Table \ref{tab:models}). As discussed below, we report posteriors only for \(5.75 \leq \mathrm{log} \frac{m}{M_\odot} \leq 9\) and \(27 \leq \frac{v}{\mathrm{km}/\mathrm{s}} \leq 450\), where the GCNN encoder distributions pass the simulation-based calibration test. Across this well-calibrated area of subhalo mass and velocity, the GCNN encoder leads to an improved figure of merit [Eq.~\eqref{eq:FoM}], which we take as the area of the posterior being reduced, i.e., tighter constraints on the subhalo parameters (up to over a factor of a hundred gain in \(FoM\)). This result is consistent with the encoder performance that we present in \S~\ref{sec:ps_gcnn_compare}, where we demonstrate that the GCNN extracts more information from the stream phase space than the power spectrum.

Both the power spectrum and GCNN encoders lead to some positive degeneracy between subhalo mass and velocity. This result is consistent with the effects of these parameters that we discuss in \S~\ref{sec:sim} and \ref{sec:PS}, where we demonstrate that the effect of a more massive subhalo can be counteracted by increasing its velocity. For the power spectrum encoder, the velocity is typically much more poorly constrained than the mass (as a fraction of the prior range). This result is consistent with the encoder performance, where we find that the fractional MAE on the velocity is always larger than for the mass.

Although the GCNN always leads to more precise posterior contours, this is particularly true as mass increases and/or velocity decreases. In these regimes, the subhalo perturbations are strongest and there is the biggest information gain from using the full phase space. For lower mass at higher velocity, there is a degree of information saturation (although the figure of merit does improve by about a factor of two and so we have not reached saturation). In this regime, the subhalo perturbations are weakest. Towards the limit of no discernible perturbation, the two encoders will tend towards the same result. The transition between these two regimes is fairly rapid; we expect this behavior from the stream simulations we show in Fig.~\ref{fig:stream_init} and the power spectra we show in Fig.~\ref{fig:ps_mass_velo}. In both the full phase space and the power spectra, we see a rapid transition in behavior above and below a mass threshold of \(\sim 10^7 M_\odot\) and a velocity threshold of \(\sim 100\) km/s, indicating that this is a physical effect manifesting in the simulations and which is correctly feeding through to the SBI. \textcolor{black}{We find that the posterior contours in Fig.~\ref{fig:post} largely follow isocontours of constant impulse, i.e., \(\frac{m}{v} = \mathrm{constant}\). This suggests that the impulse of the subhalo on the stream is the primary (though not only) quantity that determines the degeneracy between mass and velocity. We find that this approximation becomes less good as the true subhalo velocity decreases.}

\begin{figure*}
    \centering
    \includegraphics[width=0.9\textwidth]{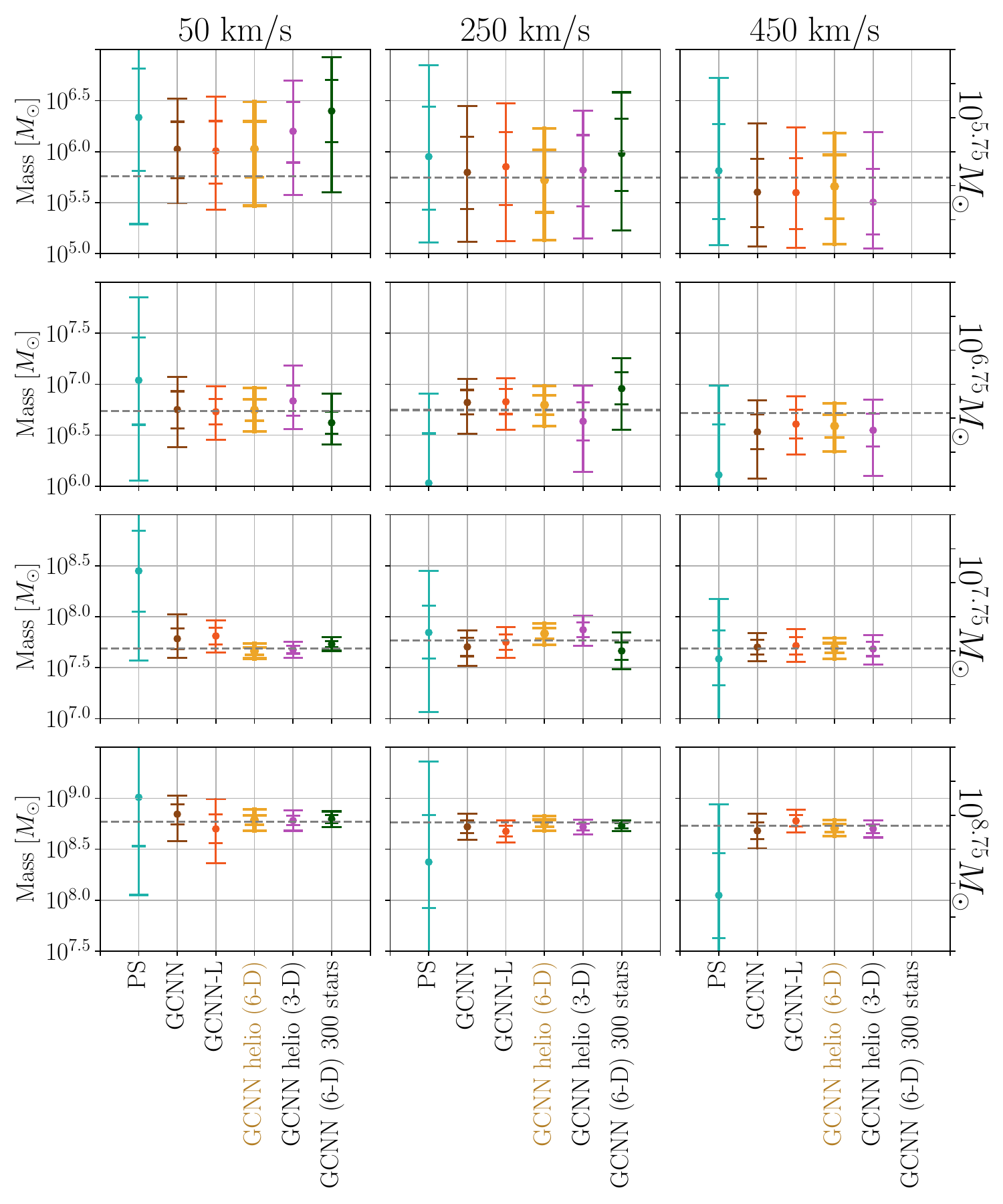}
    \caption{The marginalized posterior distribution of subhalo mass as estimated by SBI (\S~\ref{sec:sbi}) given a test set of stream simulations with true subhalo masses indicated by the grey dashed lines. \textit{In each panel}, we show subhalo mass constraints given each encoder model variant (see labels at the bottom and Table \ref{tab:models}). \textit{From top to bottom}, we vary the subhalo mass of the test simulation with the true mass given on the right hand side; \textit{from left to right}, we vary the subhalo velocity of the test simulation with the true velocity given on the top. The inner and outer ticks respectively indicate the 68\% and 95\% credible intervals, while the point indicates the mean. We do not report the GCNN (6-D) 300 stars results in the highest velocity bin as this model fails the calibration test in this regime. The GCNN helio (6-D) encoder model variant returns the strongest subhalo mass constraints.}
    \label{fig:marginalised}
\end{figure*}

With regards to determining fundamental properties of dark matter, it is most powerful to measure the masses of subhalos as these can be directly related to the subhalo mass function. In this respect, parameters of the stream perturbation, like the relative velocity between subhalo and stream, are largely nuisances (although it is conceivable that the velocity distribution of substructure could potentially be a signature of different fundamental physics). Fig.~\ref{fig:marginalised} therefore shows examples of posterior distributions of the subhalo mass, having marginalized over subhalo velocity. In this section, we consider only the two encoder model variants on the far left in each subpanel, i.e., we compare the power spectrum approach with the baseline GCNN model (see \S~\ref{sec:scale_gcnn} and \ref{sec:coord_obs} for the other encoders). The GCNN encoder always leads to stronger subhalo constraints than the power spectrum, again driven by the more informative phase space compression. The relative amount of improvement decreases as the true subhalo mass decreases, reflecting the increasing saturation in the amount of information to be extracted. The improvement is fairly independent of the true subhalo velocity.

Table \ref{tab:sbi_test} quantifies the gain in constraining power from using the GCNN encoder by evaluating the performance metric defined in Eq.~\eqref{eq:sigma_ratio} (the ratio of subhalo mass uncertainty for a given encoder over the best case; again, we consider the GCNN encoder variants in \S~\ref{sec:scale_gcnn} and \ref{sec:coord_obs}). We find that the improvement scales from a factor of five for subhalo masses \(> 10^8 M_\odot\) to 1.5 for masses \(< 10^6 M_\odot\).

\subsubsection{Simulation-based calibration test}
\label{sec:SBC}

\begin{figure}
    \includegraphics[width=\columnwidth]{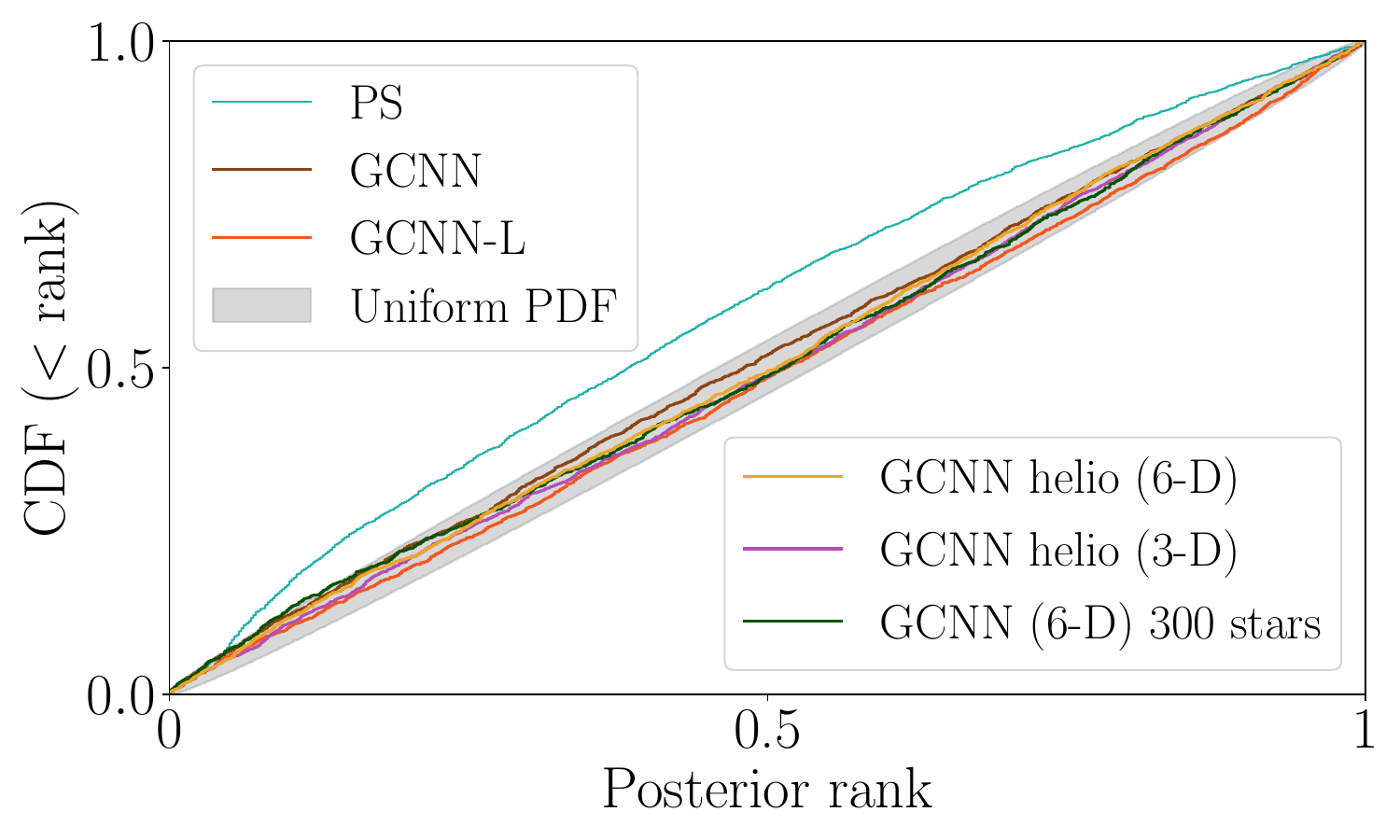}
    \caption{The cumulative distribution function (CDF) of the rank (normalized to the unit interval) of the true subhalo parameters in a set of posterior distribution samples, for a test set of stream simulations. We interpret the \(y\)-axis as the coverage probability for a given credible limit on the \(x\)-axis. As discussed in \S~\ref{sec:SBC}, we report the posterior calibration only for subhalo masses \(5.75 \leq \mathrm{log} \frac{m}{M_\odot} \leq 9\) and speeds \(27 \leq \frac{v}{\mathrm{km}/\mathrm{s}} \leq 450\), where the GCNN encoder distributions pass the simulation-based calibration test. The exception to this is the GCNN (6-D) 300 stars variant, where we report for \(27 \leq \frac{v}{\mathrm{km}/\mathrm{s}} \leq 350\). In other words, for these ranges only, the test GCNN posteriors are accurately calibrated and the rank distributions are consistent with a uniform distribution, i.e., the gray area which indicates the 99\% c.l. of a uniform probability distribution function. We find that the GCNN posteriors are not only more constraining but also more accurate than when using the power spectrum.}
    \label{fig:curve}
\end{figure}

Having discussed the information gain from using the GCNN data compression (see also \S~\ref{sec:ps_gcnn_compare}), we now assess how robust is the parameter inference by the posterior calibration tests discussed in \S~\ref{sec:metrics}. Fig.~\ref{fig:curve} shows the cumulative distribution function of the true parameter ranks within sets of posterior samples; or, the coverage probability for each credible limit. In Table \ref{tab:sbi_metric}, we report the KS \(p\) value assessing the consistency between these rank CDFs and the expected uniform CDF (gray area in Fig.~\ref{fig:curve}). Specifically, by searching through different sub-areas of the prior, we identify a common area where the GCNN encoders are well-calibrated, i.e., \(p \geq 0.05\). This well-calibrated area is for true subhalo masses \(5.75 \leq \mathrm{log} \frac{m}{M_\odot} \leq 9\) and true speeds \(27 \leq \frac{v}{\mathrm{km}/\mathrm{s}} \leq 450\), except for the GCNN (6-D) 300 stars variant, where the area is for \(27 \leq \frac{v}{\mathrm{km}/\mathrm{s}} \leq 350\) (see \S~\ref{sec:coord_obs}). It is for these areas only that we report results in Figs.~\ref{fig:post}, \ref{fig:marginalised} and \ref{fig:curve} and Tables \ref{tab:sbi_test} and \ref{tab:sbi_metric}.

We therefore find that posteriors towards the edge of the prior outside these areas are not well calibrated and therefore unreliable (i.e., when the true parameters lie outside this area). We attribute this performance to a combination of two effects. First, the edge of the prior is where the encoder performance degrades (as discussed in \S~\ref{sec:ps_gcnn_compare}). Second, there is a numerical effect where, since the posterior is cut off by the uniform prior, a discontinuity has arisen that breaks a condition for the calibration test. We discuss in \S~\ref{sec:ps_gcnn_compare} how to mitigate the encoder edge effect, while the second effect could be mitigated by using a prior with continuous support (e.g., a Gaussian) or instead learning the likelihood rather than the posterior. In any case, we find that our GCNN posteriors are well calibrated throughout the parameter space of most interest and we defer to future work a more thorough investigation of these edge effects. In particular, we find that the GCNN posteriors are always better calibrated than the power spectrum approach, meaning that using the GCNN is both more informative and more accurate. In Table \ref{tab:sbi_metric}, we also compare the negative log probabilities of the well-calibrated test set finding that the GCNN encoder model again performs better than the power spectrum encoder (lower negative probability).

\subsection{Performance scaling with number of training simulations}
\label{sec:scale_gcnn}

\begin{table}
\centering
\begin{tabular}{|c|cc|}
\hline
\textbf{Encoder model} & \textbf{Neg. log prob.} & \textbf{KS \(p\) value} \\
\hline
PS & -1.30 $\pm$ 0.90 & \(\ll 0.05\) \\
\hline
GCNN & -4.70 $\pm$ 1.61  & 0.08 \\
\hline
GCNN-L &-4.90 $\pm$ 1.62 & 0.06 \\
\hline
GCNN helio (6-D) & \textbf{-5.71 $\pm$  1.63} & 0.27 \\
\hline
GCNN helio (3-D)  & -5.08 $\pm$ 1.97 & 0.52 \\
\hline
GCNN (6-D) 300 stars &-5.40 $\pm$ 2.07 & 0.08 \\
\hline
\end{tabular}
\caption{Performance metrics (as defined in \S~\ref{sec:metrics}) on the accuracy of posterior estimation given each encoder model variant (\textit{from top to bottom}). As discussed in \S~\ref{sec:SBC}, we report the posterior calibration only for subhalo masses \(5.75 \leq \mathrm{log} \frac{m}{M_\odot} \leq 9\) and speeds \(27 \leq \frac{v}{\mathrm{km}/\mathrm{s}} \leq 450\), where the GCNN encoder distributions pass the simulation-based calibration test, i.e., \(p \geq 0.05\). The exception to this is the GCNN (6-D) 300 stars variant, where we report for \(27 \leq \frac{v}{\mathrm{km}/\mathrm{s}} \leq 350\). For the negative log probability, we give the mean and standard deviation for test samples from this well-calibrated range only. For the power spectrum approach, we report only that \(p \ll 0.05\) as it is significantly in the tail where numerical effects manifest. Posteriors are always more accurate given GCNN encoders than the power spectrum approach.}
\label{tab:sbi_metric}
\end{table}

We consider here the performance of the first of the four GCNN encoder model variants that we introduce in \S~\ref{sec:encoder} and Table \ref{tab:models}: the GCNN-L encoder. For the GCNN-L encoder, relative to the baseline GCNN encoder discussed in \S~\ref{sec:ps_gcnn_compare} and \ref{sec:SBI_calibration}, we double the number of neural network model hyper-parameters from $7 \times 10^5$ to $1.4 \times 10^6$ and, in turn, double the number of training simulations from 40,000 to 80,000. Increasing the size (two times more parameters) and depth (one additional layer) of a neural network can increase its expressivity, i.e., its ability to express complex transformations from input to output, and so we investigate if the data compression can be more informative as a consequence.

In Fig.~\ref{fig:mae_grid}, we see little systematic change in the accuracy of the encoder model from GCNN to GCNN-L; in eight out of twelve test subhalo mass and velocity bins, we see a reduced fractional error in parameter estimation. In any case, the size of the gain or loss is always much less than the gain relative to the power spectrum approach. In Fig.~\ref{fig:marginalised} and Table \ref{tab:sbi_test}, we assess the performance of the parameter inference, specifically for subhalo mass constraints. There is a discernible, though marginal, trend with up to a 12\% reduction in parameter uncertainties. We do not over-interpret this result, though note that this trend agrees with the change in encoder accuracy for subhalo masses \(> 10^6 M_\odot\). Finally, in Fig.~\ref{fig:curve} and Table \ref{tab:sbi_metric}, we report that the posteriors are well calibrated (KS \(p = 0.06\)) for the area identified in \S~\ref{sec:SBC}. The negative log probability is slightly lower than for the baseline GCNN indicating better performance. Overall, we find no strong scaling with the size of the encoder network in performance throughout the inference pipeline. For the further encoder model variants, we retain the larger architecture of the GCNN-L model.

\subsection{Performance in the limit of incomplete observations}
\label{sec:coord_obs}
We consider here the performance of the final three GCNN encoder model variants that we introduce in \S~\ref{sec:encoder} and Table \ref{tab:models}: the GCNN helio (6-D), GCNN helio (3-D) and GCNN (6-D) 300 stars encoders. For the GCNN helio (6-D) encoder, relative to the baseline GCNN and the GCNN-L encoders, we change the input coordinate system of the stream phase space. We change the coordinates from galactocentric spherical (i.e., the radius, polar angle and azimuthal angle relative to the center of the Galaxy and their velocity equivalents) to a set of heliocentric coordinates (RA, DEC, heliocentric distance, \(v_\mathrm{RA}, v_\mathrm{DEC}, v_\mathrm{LOS}\)). In doing this, we simulate more closely to real data, where we observe in angular coordinates on the sky, measure line of sight velocities from stellar spectra, infer heliocentric distances from stellar parallaxes and measure angular velocities indirectly though proper motions. We convert proper motions into angular velocities in order to decorrelate the position and velocity vectors as input features to the encoder. We may expect some sensitivity of performance to the change of coordinate system, despite no change in input information content, as discussed in \S~\ref{sec:coord_discussion}. We quantitatively investigate this effect here.

In Fig.~\ref{fig:mae_grid}, we see that the GCNN helio (6-D) encoder performs more accurate parameter estimation than both the GCNN and GCNN-L models in ten out of twelve test subhalo mass and velocity bins. This result feeds through to stronger subhalo mass constraints after parameter inference (Fig.~\ref{fig:marginalised} and Table \ref{tab:sbi_test}). We find that the improvement (relative to GCNN-L) scales from a factor of 1.15 for subhalo masses \(< 10^6 M_\odot\) up to a factor of 1.83 for masses \(> 10^8 M_\odot\). The fact that the gain is largest for the most disruptive subhalo interactions suggests that it is the projection of clear features in the stream like gaps and spurs that drives this result. Indeed, Fig.~\ref{fig:stream_init} demonstrates that features like spurs are more discernible in certain coordinates (for the simulations in the figure, \(\phi_2\) and heliocentric distance rather than \(\phi_1\), RA and DEC). It follows that changing from galactocentric to heliocentric coordinates will change the prominence of certain stream features. Doing this transformation in turn changes the feature space which is input to the encoder model, thus making an informative data compression an easier or harder task. We discuss in \S~\ref{sec:coord_discussion} other examples where the feature embedding space affects model performance. Fig.~\ref{fig:curve} and Table \ref{tab:sbi_metric} again confirm that the posteriors are well-calibrated in the area we have identified (\(p = 0.27\), see \S~\ref{sec:SBC}). The negative log probability is even lower than for the GCNN and GCNN-L models, indicating further improved performance.

With the GCNN helio (3-D) and GCNN (6-D) 300 stars encoders, we probe how the performance of the inference pipeline scales in the limit of a reduced set of observations. For the other GCNN encoders, we allow access to the full 6-D phase space of 3000 {\gdone} stars. This setting is conceivable through a future combination of deep photometric, spectroscopic and astrometric observations, but is beyond our current capability. For the GCNN helio (3-D) encoder, we input 3000 stars but only with a 3-D phase space, which we take as RA, DEC and \(v_\mathrm{LOS}\). This setting roughly mocks up a future deep photometric survey with follow-up spectroscopy. The number of data is reduced from 18,000 to 9000. For the GCNN (6-D) 300 stars encoder, we input only 300 stars but with a full 6-D phase space. This setting roughly mocks up observations we will have in the near future, where we will have complete data for a subset of the {\gdone} stars. The number of data here is reduced further to 1800. We do not set out to perform detailed forecasts for particular observational settings, but rather we seek to understand the trade-off in performance from having more stars with incomplete data or fewer stars with complete data. We leave for future work a detailed forecast using our new data analysis method.

In Fig.~\ref{fig:mae_grid}, we see that the GCNN encoders with reduced input perform less precise parameter estimation than the GCNN helio (6-D) model in all but one of the twelve test subhalo mass and velocity bins. This result is expected as supplying less information to the model will lead to a less informative summary statistic. However, both the GCNN helio (3-D) model with 9000 input data and the GCNN (6-D) 300 stars model with 1800 input data perform better at parameter estimation than the power spectrum approach (with 3000 input data before any compression, i.e., the \(\phi_1\) coordinates of 3000 stars). This result underscores the combined benefit of a more informative compression scheme (GCNN over PS) and gathering multi-dimensional data (more than 1-D). Between the two reduced input models, we find that the GCNN helio (3-D) performs more accurate parameter estimation in nine out of twelve parameter bins.

In Fig.~\ref{fig:marginalised} and Table \ref{tab:sbi_test}, we find that the scaling of the constraining power of the parameter inference is consistent with the scaling of the information extracted in the data compression. The \(1 \sigma\) uncertainty on subhalo mass is worse than when using all the input data across the prior range that we consider (apart from one of eight test subhalo mass bins), but always better than when using the power spectrum pre-compression, even if there are fewer input data (1800 for the GCNN (6-D) 300 stars model, 3000 for the PS model). The degradation from using incomplete observations relative to the GCNN helio (6-D) model is strongest for intermediate subhalo masses \(\sim 10^7 M_\odot\). We see the effects of information saturation at lower masses, where all models perform increasingly similarly as the subhalo perturbation weakens. For the most massive subhalos \(\sim 10^8 M_\odot\), we even see a marginal gain in constraining power from inputting fewer data (the GCNN (6-D) 300 stars model). This result tells us that, when the perturbation is very strong, there are a few stars that dominate the signal. Therefore, adding more unperturbed stars as input can make this signature less clear in the feature embedding space.

Finally, in Fig.~\ref{fig:curve} and Table \ref{tab:sbi_metric}, for the GCNN helio (3-D) variant, we report well-calibrated posteriors (\(p = 0.52\)) in the range for subhalo masses \(5.75 \leq \mathrm{log} \frac{m}{M_\odot} \leq 9\) and speeds \(27 \leq \frac{v}{\mathrm{km}/\mathrm{s}} \leq 450\). For the GCNN (6-D) 300 stars variant, we report a slightly reduced well-calibrated range (\(p = 0.08\)) for \(27 \leq \frac{v}{\mathrm{km}/\mathrm{s}} \leq 350\), suggesting that the reduced amount of data can make the SBI task harder. The reduced amount of data in both variants leads to worse negative log probabilities, albeit still better than the power spectrum encoder.

\section{Discussion}
\label{sec:discussion}
\subsection{Improved model parameter estimators}
\label{sec:disc}
In \S~\ref{sec:results}, we present results demonstrating that the GCNN encoders return more accurate determinations of the perturbing subhalo mass and speed than when using the power spectrum. This information gain persists, irrespective of the size of the GCNN model, the coordinate system of the data or the subset of the data that is input to the GCNN.

We discuss here what drives this gain in information being extracted from the stream data. First, the GCNN ingests a richer dataset since it includes the full 6-D phase space. On the other hand, the power spectrum model uses only the perturbations along the $\phi_1$ coordinate, which is a lossy compression. Second, the dataset is inherently a particle-like system. Thus, there are symmetries underlying the system like equivariance of which the GCNN takes full advantage due to the bidirectional message passing architecture. This symmetry is not used in the power spectrum apparoach since the inputs are compressed by binning. Further, the binning preprocessing for the power spectrum also smoothes out potential signals in the data. Lastly, the GCNN model makes fewer assumptions about the data. Current empirical evidence in the trends of deep learning has seen an increase in abstraction of the features in the data. This means that the less hard coded is the feature extraction, the better the model can perform. The improved flexibility of the GCNN to adapt to a given problem often yields better results at the cost of computational expense and some loss of interpretability, which we observe in our experiments.

\subsection{Breaking model parameter degeneracies}
As discussed in \S~\ref{sec:sim}, there is a physical degeneracy in the stream simulations between increasing subhalo mass and speed. As the subhalo becomes more massive, there is more stream perturbation, but this effect can be counteracted by increasing the relative velocity between subhalo and stream so that the impulse is reduced. In \S~\ref{sec:SBI_calibration}, we see this degeneracy manifest in the posterior distributions whether using the power spectrum or GCNN encoders. A striking result is that for the power spectrum model, the subhalo velocity is largely unconstrained and independent of the true parameters. This lack of sensitivity is consistent with the smaller dynamic range in power spectra that we see in Fig.~\ref{fig:ps_mass_velo} as we vary subhalo velocity compared to mass. In Fig.~\ref{fig:post}, we see that using the GCNN encoder can significantly break this degeneracy in a mass- and velocity-dependent way. As the mass decreases and/or the velocity decreases, the GCNN model returns much stronger constraints on the subhalo velocity, thus (partially) breaking the degeneracy with subhalo mass. We attribute this behavior to the fact that the GCNN has additional input stream velocity information, which is more sensitive to the subhalo velocity than the power spectrum. This result emphasises the importance of providing stream star velocities in constraining the mass of perturbing subhalos.

\subsection{Saturation in GCNN performance}
In \S~\ref{sec:scale_gcnn}, we find that increasing the size and depth of the GCNN model leads to marginal gain in parameter estimation and the precision and accuracy of posterior estimation. This result indicates that we can likely achieve similar performance with fewer training simulations than previously anticipated. This can be attributed to a known property of over-smoothing \citep{smoothing} of the GCNN, where large graph models [number of hyperparameters $\sim \mathcal{O}\left(10^6\right)$] have plateauing performance as the number of hyperparameters increases with more layers \citep{graph_scale}. Thus, if increasing model capacity cannot yield substantial improvements, it is unsurprising that increasing the amount of training data also does not. For this problem, we suggest that $\mathcal{O}(10^3 - 10^4)$ training simulations thus suffices while remaining computationally efficient. Further, in future work, we will investigate more efficient use of training data through sequential SBI methods that iteratively generate training data in a more optimal fashion.

\subsection{Effect of the stream coordinate system}
\label{sec:coord_discussion}
In \S~\ref{sec:coord_obs}, we find up to a factor of 1.83 improvement in the subhalo mass constraint from changing from galactocentric to heliocentric coordinates. This is initially surprising as an invertible coordinate transformation does not directly yield any additional information about the dynamics of the system, yet performance improves. This phenomenon is well studied in the field of representation learning \citep{ML_textbook_Goodfellow}. The choice of embedding or representation of data is central in many deep learning models. E.g., positional encoding, which is an invertible map to represent position data using sinusoids, when applied to physics inspired neural networks \citep{pos_enc, pinns}, improves the learning outcomes. In natural language processing, tokenization using \textsc{word2vec} \citep{token} or word embeddings in transformer models  \citep{attention} adopt a ``basis change" to boost model performance. Although it appears surprising at first that a coordinate change can yield noticeable model improvements, this is a norm rather than an exception. The common argument is that the choice of coordinates can disentangle features. It is arguable that stream perturbations are more aligned along a particular axis in the heliocentric coordinates than the galactocentric coordinates. Indeed, we see in Fig.~\ref{fig:stream_init} that, for these example simulations, the spur is more or less discernible depending on the coordinates in which it is projected. For real data, it will therefore be critical to assess the sensitivity of results to the chosen coordinates.

\subsection{Future simulation improvements}
\label{sec:discussion_sims}

While this work demonstrates a proof-of-concept method for characterizing dark matter subhalo perturbations in stellar streams, future iterations will incorporate more sophisticated simulation methods with a view to a robust data analysis. First, we will vary additional simulation parameters such as the impact parameter, the angle of approach of the subhalo to the stream, the time of closest approach and the subhalo density profile and size. It is important to consider these parameters in addition as their values are unknown for any given stream and so we must constrain them using data. Second, we will incorporate multiple subhalo interactions. The host (MW) halo contains a distribution of subhalos (of varying mass, velocity, etc.) and typically tens or hundreds of these will appreciably interact with a stream. Thus, folding in a realistic subhalo distribution and multiple interactions will be critical for a future comparison to data. We also anticipate investigating more realistic particle-spray simulation approaches \citep{Carlberg2018,Malhan2019,mass,Qian2022}, modeling time-dependent variations in the background potential \citep{Buist2015,Koppelman2021,Brooks2024} and including baryonic perturbers like giant molecular clouds \citep{jo}. Finally, for a reliable comparison to data, we must model the properties of the survey(s) when generating the input simulations. E.g., we can model measurement uncertainties by sampling from a distribution that resembles instrument uncertainties and we can further account for completeness and masking by post-processing the simulations before input to the encoder.

\section{Conclusions}
\label{sec:concs}

We have demonstrated that the combination of a graph convolutional neural network compression and simulation-based inference can improve constraints on the mass of a perturbing subhalo by factors of three to eleven compared to the current state-of-the-art density power spectrum analysis of a GD-1-like stellar stream. When varying the mass and velocity of the subhalo in our simulations, we find that the GCNN encoder consistently improves estimation of these parameters relative to the power spectrum. The ability of the GCNN to exploit information in the kinematics of the stream improves inference of the subhalo velocity, which in turn breaks the degeneracy with subhalo mass. We find that posterior distributions given the GCNN compression are significantly better calibrated, meaning that this method is simultaneously more precise and more accurate.

We perform three experiments with the GCNN encoder. First, we find that there is no strong scaling in the performance of the method as we increase the complexity of the neural network and increase the number of training simulations. We conclude that \(\mathcal{O}(10^3 - 10^4)\) stream simulations are sufficient in training compression networks for future analyses, thus informing the computational expense of the problem. Second, we find some sensitivity in our results to the coordinate system of the stream as input to the pipeline. This result is consistent with the deep learning literature and we postulate that identifying a projection of the stream that accentuates the features associated with a subhalo interaction can enhance the ability of the encoder to extract information.

Finally, while we do not perform detailed forecasts for the power of this method applied to upcoming data with a full treatment of errors and systematic effects in this proof-of-principle study, we approximate two observational settings achievable in the near future. We compare a setting with 3000 stars with sky coordinates and line-of-sight velocities (a combination of deep photometry and spectroscopy but without associated astrometry) with a setting in which the dataset consists of 300 stars with full 6-D phase space data (photometry, spectroscopy and astrometry). We find an improvement in subhalo mass constraints by factors of two to eleven for masses from \(\left(10^6 - 10^9\right)M_\odot\) compared to the constraints when using the power spectrum approach. The improvement is about the same for the two settings suggesting that collecting exhaustive information on a few hundred stream stars can be about as powerful as a deeper survey of thousands of members with only RA, DEC and line-of-sight velocities. We conclude that the graph neural network method we have introduced here will be powerful in maximizing the information return about the MW subhalo population from upcoming deep- and wide-field photometric and spectroscopic surveys (e.g., \textit{Rubin}, \textit{Euclid}, DESI, \textit{Roman}). In future work, we will develop this method further to incorporate more sophisticated simulation approaches and the modeling of survey properties with a view to future data analysis.

\begin{acknowledgments}
\section{Acknowledgments}
The authors thank Jo Bovy for insightful discussions and support with using \textsc{galpy}. KKR is supported by an Ernest Rutherford Fellowship from the UKRI Science and Technology Facilities Council (grant number ST/Z510191/1). The Dunlap Institute is funded through an endowment established by the David Dunlap family and the University of Toronto. TSL acknowledges financial support from the Natural Sciences and Engineering Research Council (NSERC) of Canada through grant RGPIN-2022-04794. RH acknowledges support from the Natural Sciences and Engineering Research Council of Canada Discovery Grant Program and the Connaught Fund. The authors at the University of Toronto acknowledge that the land on which the University of Toronto is built is the traditional territory of the Haudenosaunee and, most recently, the territory of the Mississaugas of the New Credit First Nation. They are grateful to have the opportunity to work in the community on this territory. PXM thanks the friends at CERN for the generous vehicular support and fantastic company at the early stages of this work.
\end{acknowledgments}

\appendix

\section{Neural architecture optimization}
\label{sec:hyperparam}
We optimize the architecture of the power spectrum and baseline GCNN compression networks using the \texttt{scikit-optimize} Bayesian optimization \citep{baeys_tune} package\footnote{\url{https://scikit-optimize.github.io/stable}.}. We search for the architecture that minimizes the validation loss [Eq.~\eqref{eq:mae}]. Table \ref{tab:4model_param} shows the architecture parameters that we vary and the values that we consider. For this search, we train with only 1000 simulations and the \textsc{ADAM} optimizer for 200 epochs. The bold values in Table \ref{tab:4model_param} are the optimal choice given this procedure.

\begin{table}
    \centering
   \color{black}\begin{tabular}{p{6cm}c}
  \hline
\multicolumn{2}{c}{\textbf{Power spectrum encoder}}\\
\hline
Number of dense layers \# 1 & 1, 2, \textbf{3}\\
Number of dense layers \# 2 & 1, 2, \textbf{3}\\
Number of dense layers \# 3 & 1, 2, \textbf{3}\\
Dense layer \# 1 width & 256, \textbf{512}, 1024, 2048\\
Dense layer \# 2 width & 64, 128, \textbf{256}\\
Dense layer \# 3 width & \textbf{16}, 32, 64\\
Activation function & \textbf{Leaky ReLU}, ReLU, Sigmoid, Tanh\\
\hline
\hline
\multicolumn{2}{c}{\textbf{Graph convolutional neural network (GCNN) encoder}} \\
\hline
Number of graph convolutional layers & 1, 2, \textbf{3}, 4, 5\\
Graph convolutional layer width & 6, 32, 64, \textbf{128}, 512\\
Number of dense layers & 1, \textbf{2}, 3, 4, 5\\
Dense layer width & 32, 64, 128, \textbf{512}, 1024\\
Activation function & Leaky ReLU, \textbf{ReLU}, Sigmoid, Tanh\\
\hline
\end{tabular}
\caption{Results of the encoder neural architecture search using Bayesian optimization. The \textit{top} part shows results for the power spectrum encoder, while the \textit{bottom} part shows results for the baseline GCNN encoder. Each of the parameters on the \textit{left} is optimized while searching across the values on the \textit{right}. The optimal architectures are indicated in \textit{bold}. \label{tab:4model_param}}
\end{table}

\section{Training and validation loss curves}
\label{sec:training}
\begin{figure*}
    \centering
    \begin{subfigure}[b]{0.49\textwidth}
        \centering
        \includegraphics[width=\textwidth]{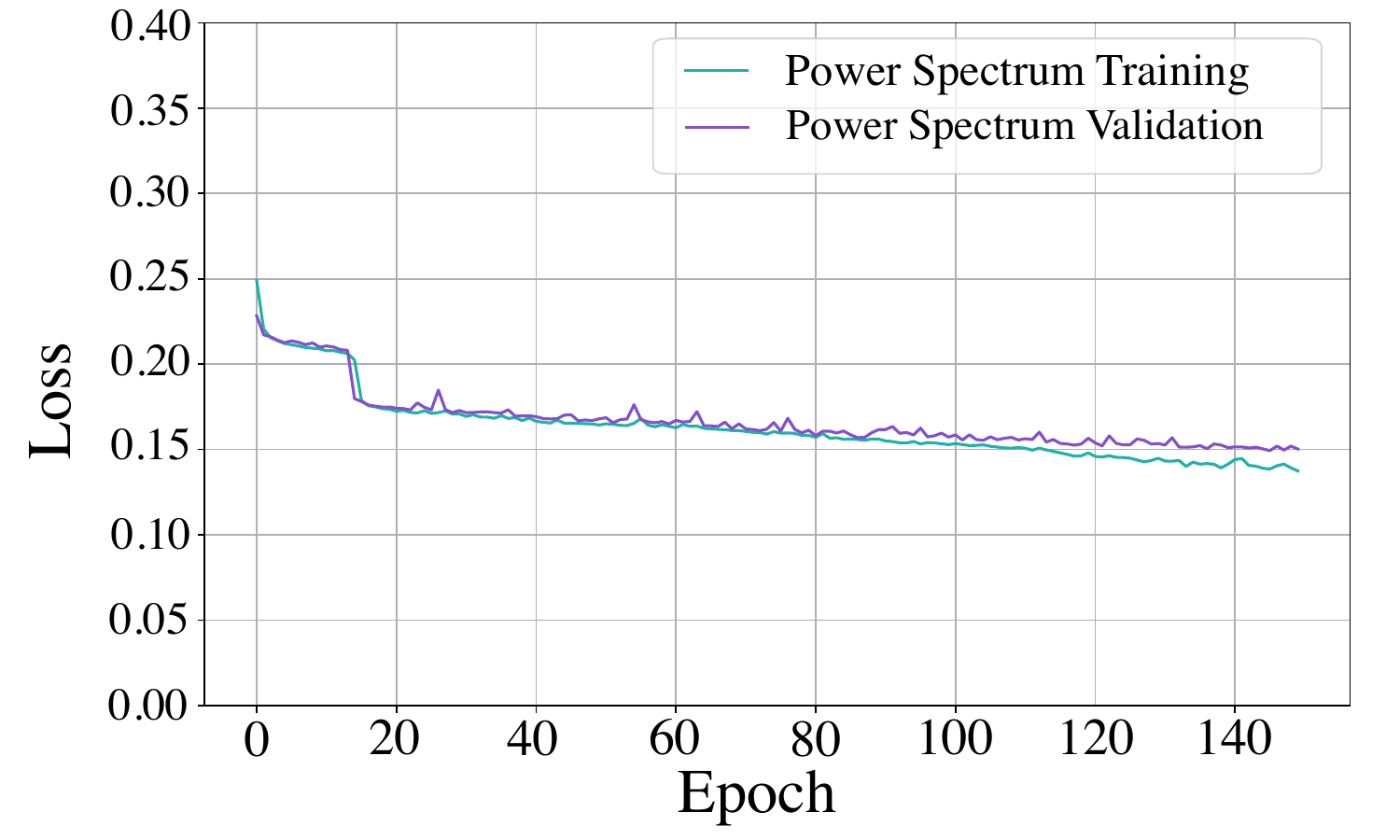}
    \end{subfigure}
    \begin{subfigure}[b]{0.49\textwidth}
        \centering
        \includegraphics[width=\textwidth]{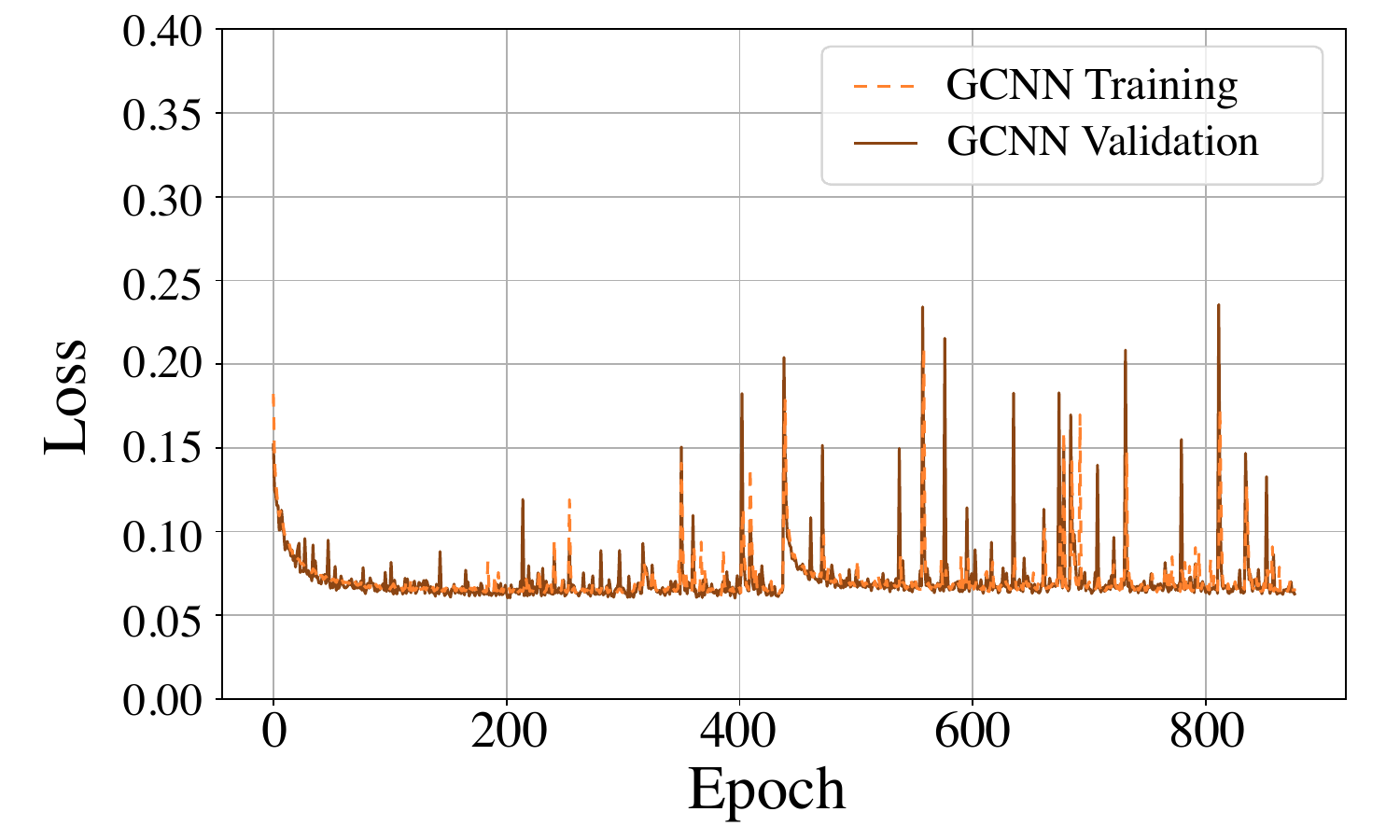}
    \end{subfigure}
    \caption{The training and validation loss curves as a function of epoch for the power spectrum (\textit{left}) and baseline GCNN (\textit{right}) encoders. The GCNN training returns a lower final loss, while the training and validation losses do not diverge indicating that overfitting has been avoided.}
    \label{fig:loss}
\end{figure*}

Figure \ref{fig:loss} shows the training and validation losses [Eq.~\eqref{eq:mae}] as a function of epoch for the power spectrum and baseline GCNN encoders. The GCNN encoder finishes with a lower loss indicating a better fit, while the training and validation curves do not diverge indicating that overfitting to the training set has not occurred.

\section{Visualization of a graph convolutional layer}
\label{sec:visualise}
\begin{figure*}
  \centering
  \includegraphics[width=1\linewidth]{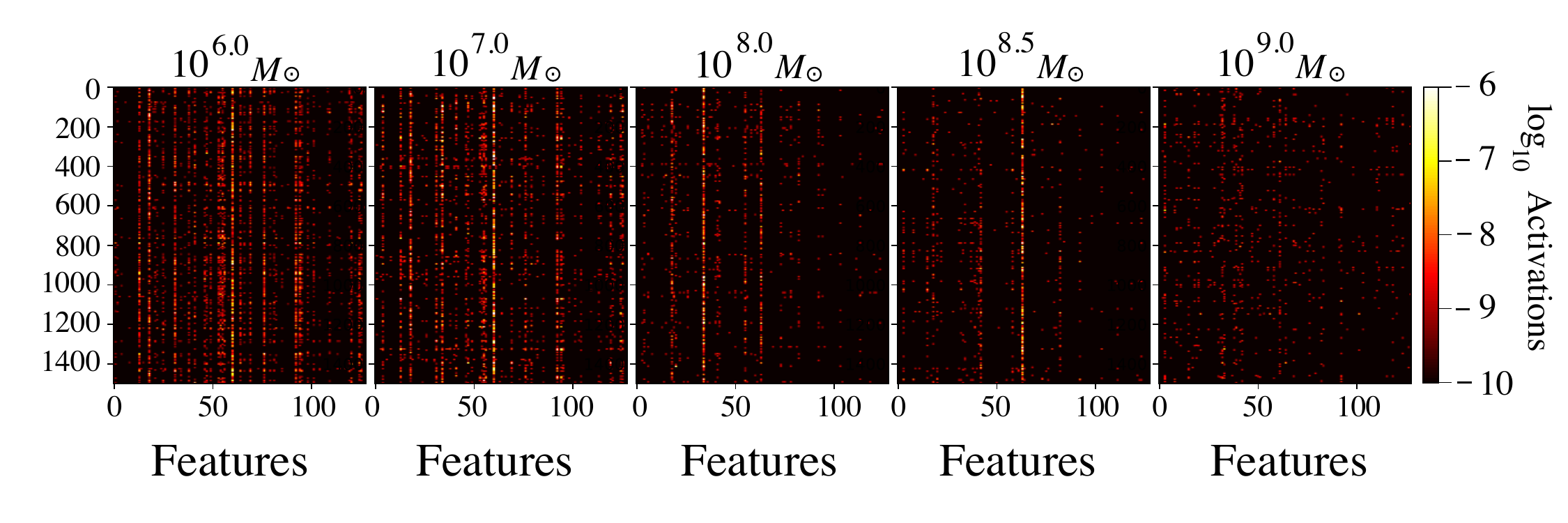}
  \caption{The activation in the final graph convolutional layer of the baseline GCNN model for one arm of the stellar stream (dimensions of 1500 stars (nodes) by 128 features). \textit{From left to right}, as indicated at the top, we vary the mass of the perturbing subhalo in the input simulation but otherwise fix all parameters including the random seed and the velocity of the subhalo relative to the stream to 440 km/s. As the mass varies, different features are activated in the network.}
  \label{fig:latent_space}
\end{figure*}

The GCNN approach to analyzing stellar streams that we have introduced, although more powerful, is necessarily more black box than measuring the density power spectrum. Nonetheless, we can open up the black box in order to increase interpretability and to inform better network construction. Fig.~\ref{fig:latent_space} visualizes the final graph convolutional layer in the baseline GCNN model. This is the final layer before the aggregation across nodes. We perform an experiment where we fix the GCNN model and then vary the input to the network by changing only the mass of the perturbing subhalo (random seed also fixed). In this way, we see which features are activated by our parameter of interest. The latent space remains sparse, i.e., only a few features are significantly activated. As the mass increases, an increasingly small number of nodes (stars) and features are significantly activated supporting the success of the GCNN (6-D) 300 stars variant in this regime.

\bibliography{main}{}
\bibliographystyle{aasjournal}

\end{document}